\documentclass[twocolumn,superscriptaddress,aps,prb,longbibliography]{revtex4-2}
\usepackage[utf8]{inputenc}
\usepackage[T1]{fontenc}
\usepackage{newtxtext,newtxmath}
\usepackage{graphicx}
\usepackage[colorlinks]{hyperref}
\usepackage{bm}
\usepackage{microtype}
\usepackage{physics}
\usepackage{xcolor}
\usepackage[caption=false]{subfig}
\usepackage{orcidlink}
\usepackage{relsize}

\newcommand{\xopr}[3][]{X_{#1}^{#2 #3}}
\newcommand{\taus}[1][]{\tau_{#1}^{\sigma \sigma'}}

\begin{document}
\title{Re-entrant topological order in strongly correlated nanowire due to Rashba spin-orbit coupling}

\author{Kaushal Kumar Kesharpu\orcidlink{0000-0003-4933-6819}}
\email{kesharpu@theor.jinr.ru}

\author{Evgenii A. Kochetov}
\affiliation{Bogoliubov Laboratory of Theoretical Physics, Joint Institute for Nuclear Research, Dubna, Moscow Region 141980, Russia}

\author{Alvaro Ferraz}
\affiliation{International Institute of Physics - UFRN, Department of Experimental and Theoretical Physics - UFRN, Natal 59078-970, Brazil}

\date{\today}

\begin{abstract}
  The effect of the Rashba spin orbit coupling (RSOC) on the topological properties of the one-dimensional (1D) extended \emph{s}-wave superconducting Hamiltonian, in the presence of strong electron-electron correlation, is investigated. It is found that a non-zero RSOC increases the periodicity of the effective Hamiltonian, which results in the folding of the Brillouin zone (BZ), and consequently in the emergence of an energy gap at the boundary of the BZ. Essentially the initial single band is divided into number of sub-bands. If the chemical potential lies inside the energy gaps (sub-bands) then the phase is topologically trivial (non trivial). This is the origin of re-entrant nature of the existent topological properties. The emergence of sub-bands allows us to drive the system in and out of the topological phase by the proper tuning of the chemical potential. A heterostructure involving van der Waals materials and a 1D Moire pattern for an investigation of the predicted effect has also been proposed. We also discuss how in-plane magnetic field can be used to control the RSOC coupling and induced periodicity in depleted InAs nanowire in which evidence of strong electron-electron correlation has been found.
\end{abstract}

\maketitle

\section{Introduction}
\label{sec:introduction}

\begin{figure}[tbh]
  \centering
  \subfloat[][]{\includegraphics[width=0.48\textwidth]{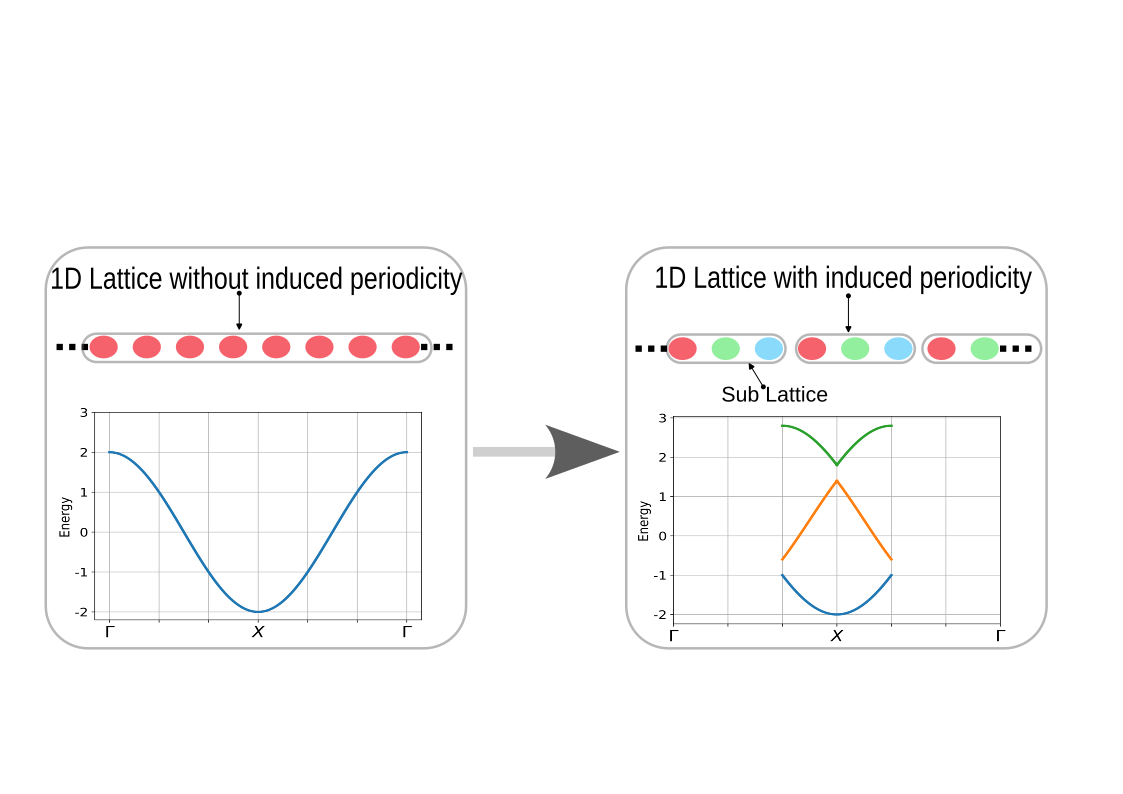}\label{fig:scheme-folding-of-BZ}}\\
  \subfloat[][]{\includegraphics[width=0.45\textwidth]{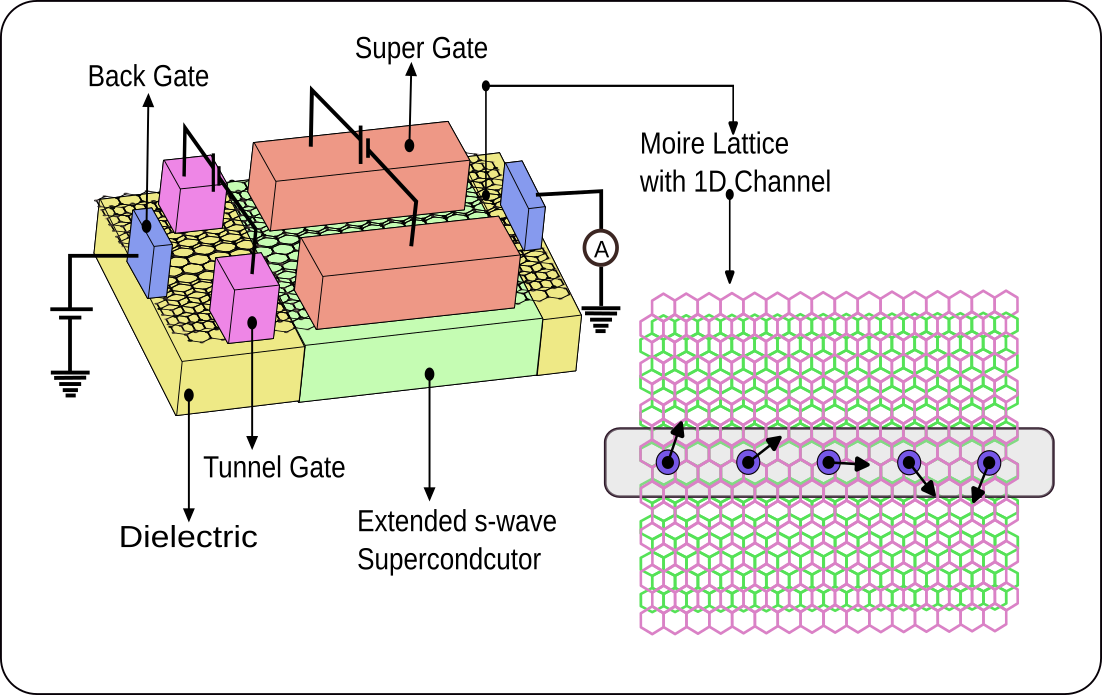}\label{fig:scheme-device-moire}}\\
  \caption{(a) The schematics of the energy dispersion of the non-interacting electrons in a 1D chain with (right) and without (left) a three fold periodicity of the lattice. The three fold periodicity results in the folding of the energy dispersion three times and, consequently, in the emergence of the energy gap at the BZ boundaries. If the chemical potential lies within this gap and no other band is involved no topological phase is present. On the right hand side diagram, the  tuning of the chemical potential from the lower to the higher band will result in the re-entrant topological order. However, this is absent on the left hand side diagram, as there is no energy gap. (b) Schematics of the proposed heterostructure involving (quasi-)1D Moire lattice for detection of the re-entrant topological phase in strongly correlated system (discussed in Sec. \ref{sec:heter-exper-real}).}
  \label{fig:schematics}
\end{figure}

The prediction of spatially separated Majorana fermions (MF) in a 1D topological superconductor (TSC) \cite{kitaev_2001_UnpairedMajorana_Phys-Usp} has marked the paradigm shift in the research for the realization of quantum computers \cite{sarma_2015_MajoranaZero_NpjQuantumInf}. Several experimental platforms have been proposed to realize MF \cite{yazdani-2023-huntin-major,flensberg-2021-engin-platf}; among them heterostructures of a semiconducting nanowire (NW) on top of a $s$-wave SC have got a special attention, mainly due to the ease of synthesis and widespread availability of the constituent materials \cite{lutchyn_2018_MajoranaZero_NatRevMater}. Three ingredients are necessary for such NW/SC heterostructures: (i) the RSOC ($\alpha$), (ii) an external magnetic field ($B_{z}$), (iii) the proximity induced s-wave superconducting (SC) order parameter ($\Delta$). The RSOC shifts the momentum parabola sideways locking the electron spin with momentum ($k$), $B_{z}$ opens a gap at $k=0$, and $\Delta$ opens a gap at opposite Fermi points.

Experimental efforts involving NW/SC have been partially fruitful and are still ongoing \cite{mandal-2023-topol-super,flensberg-2021-engin-platf,frolov-2020-topol-super}. Some of the major concerns regarding NW/SC heterostructures are as follows. The critical field ($H_{c2}$) and the critical temperature ($T_{c}$) of the usual $s$-wave SC are small. Low $H_{c2}$ requires a large Lande $g$ factor in a nanowire so that large gap can be opened at $k=0$ under $B_{z}<H_{c2}$. A large $\alpha$ in NW is desirable as it allows for stronger spin and momentum locking. From the design point of view, the presence of a magnetic field in electronic circuits is far from ideal, as stray fields hamper the miniaturization of the circuit. The interface between the NW and the SC should be both of high quality (allowing for a higher proximity induced SC gap) and smooth (to suppress disorder induced scattering which is detrimental for proximity induced SC). Although, the latter concerns are of a more applied nature, however, the former (i.e., the large $\alpha$, $g$) is more of fundamental. Hence, the question arises, whether the requirements for having both RSOC and magnetic field can be completely avoided?

A number of alternative platforms have been proposed which don't require neither RSOC nor magnetic field. Some of them are associated with atop of a  magnetic atomic chain on a SC \cite{pawlak-2019-major-fermion}, topological insulator/magnetic insulator/SC heterostructure \cite{fu_2008_SuperconductingProximity_PhysRevLett}, monolayers of WTe$_{2}$ \cite{fatemi-2018-elect-tunab}, van der Waals (vdW) materials \cite{you-2021-two-dimen,li-2021-obser-topol,kezilebieke-2022-moire-enabl}. Another interesting proposal is to use strong electron-electron (\emph{e-e}) correlation, in place of a magnetic field, to access the topological SC phase \cite{stoudenmire_2011_InteractionEffects_PhysRevB,aksenov_2020_StrongCoulomb_PhysRevB,aksenov-2023-effec-local,zlotnikov-2020-spin-orbit,kesharpu-2024-propos-realiz}. Using DMRG it was found that strong \emph{e-e} correlation favors the formation of the MF even without magnetic field \cite{stoudenmire_2011_InteractionEffects_PhysRevB,aksenov_2020_StrongCoulomb_PhysRevB}. This favorable scenario can be easily understood in the context of  a 1D chain with a single valence electron at every atomic site. Under a magnetic field the spin of every electron is frozen along the direction of the applied field. This results in every atomic site containing at most a single electron, the so called "no double occupancy" (NDO) condition. The NDO scenario can also appear due to strong \emph{e-e} correlation, as the energy penalty for placing one more electron at an already occupied atomic site will be too high for that to take place; of course there is a caveat: the spin is not automatically frozen along a single direction in this case. One way to solve this problem is to invoke the so-so called "spin-charge separation" method \cite{giamarchi_2004_QuantumPhysics_}. Due to the strong \emph{e-e} interaction the electron-hole excitations delocalize the electron until it becomes totally incoherent. As a result at low energy for this strong coupling regime, there remains only separate non interacting collective spin (spinon)  and charge (holon) excitations.

Although \emph{e-e} correlation is an intrinsic property of the material, but several recent proposals have been put forward to control it extrinsically \cite{sato-2019-stron-elect,kim-2020-contr-elect,balents-2020-super-stron}. With these considerations, in this article we investigate the Hamiltonian of a (quasi-)1D topological SC system in the presence of strong \emph{e-e} correlation. The SC order parameter is of extended \emph{s}-wave \footnote{Superconducting order parameter $\Delta$ can be either $s$-wave or extended $s$-wave depending on its symmetry in momentum space. For $s$-wave the $\Delta$ is isotropic and has $C_{4}$ symmetry, however, for extended $s$-wave $\Delta$ is anisotropic and has $C_{4}$ symmetry \cite{sigrist-1991-phenom-theor}.}; proximity induced due to iron based SC. The reason for choosing iron-based SC is because they have higher $T_{c}$, $H_{c2}$ and superconducting gap \cite{hirschfeld-2015-using-gap,hosono_2018_RecentAdvances_MaterialsToday}. In the Hamiltonian we also included RSOC and magnetic field to understand their affect on the topological properties; although, the overall goal is to get rid of them. To solve the Hamiltonian we use spin-charge separation and $su(2|1)$ path integral method \cite{ferraz_2011_EffectiveAction_NuclearPhysicsB}. This method has been applied to several other systems \cite{kesharpu_2024_TopologicalHall_PhysRevB,kesharpu_2023_TopologicalHall_PhysRevB,kesharpu-2024-propos-realiz,ferraz_2023_ConnectionKitaev_AnnalsOfPhysics,kesharpu-2024-treat-stron,ferraz_2022_FractionalizationStrongly_PhysRevB}.

The main results of this work is as follows. (i) For non-zero RSOC the system can enter into the topological phase multiple times when the parameters are appropriately tuned (see Figs. \ref{fig:seq-phase-diag-new} and \ref{fig:seq-phase-diag}). This happens due to the folding of the Brillouin zone (BZ), which takes place because of the quasi-periodicity induced by the non-zero RSOC as shown schematically in Fig. \ref{fig:scheme-folding-of-BZ}. The same property was also observed in Ref. \cite{kezilebieke-2022-moire-enabl} due to Moire potential. In Ref. \cite{heedt-2017-signat-inter} a related effect, re-entrant conductance, in depleted InAs NW due to RSOC and strong \emph{e-e} correlation was also observed. (ii) One can use only chemical potential as the tuning parameter to drive the system into and out of the topological phase (see Fig. \ref{fig:seq-dynamic-theta}). Basically, the effective Hamiltonian, Eq. (\ref{eq:effectiv-ham-spiral-spin}), contains the dynamical parameter $\theta$ (in the sense it depends on other system paramters), which loosely speaking corresponds to the modulation angle of spiral spin field. For system to be in the topological phase the two conditions should be satisfied, firstly the $\theta$ should be in the range $\theta \in \left( 0,\pi \right)$, secondly the $\mathbb{Z}_{2}$ invariant of the Hamiltonian should be $Q=-1$. From the free energy calculation we find that both these conditions can be satisfied in some defined range of chemical potential. (iii) We propose heterostructure involving vdW materials for experimental realization of the effect; a schematic diagram of the same is shown in Fig. \ref{fig:scheme-device-moire}. We discuss about this heterostructure thoroughly in Sec. \ref{sec:heter-exper-real}. (iv) In Sec. \ref{sec:contr-peri-hamilt} we investigate the effect of in-plane magnetic field on the Hamiltonian. A heterostructure (see Fig. \ref{fig:scheme-InAs}) involving depleted InAs NW under external magnetic field and electric field is proposed for observation of the discussed effects.

\section{Model}
\label{sec:model}

We start from a Hubbard type Hamiltonian of a 1D SC NW under an externally applied magnetic field with Rashba spin-orbit coupling (RSOC):
\begin{equation}
  \label{eq:basic-ham}
  \begin{aligned}
    H=&-c^{\dagger}_{i\sigma} \left( t \taus[0] + \alpha \taus[y] \right) c_{i+1\sigma'}
        -\imath c^{\dagger}_{i\sigma}\Delta \taus[y]c^{\dagger}_{i+1 \sigma'}+\text{H.c.}\\
      &\quad + c^{\dagger}_{i\sigma} \left( \mu \taus[0] + B_{z} \taus[z] \right) c_{i+1\sigma'}+ U\sum_in_{i\uparrow}n_{i\downarrow}.
  \end{aligned}
\end{equation}
Here, $c_{i\sigma}^{\dagger}$ ($c_{i\sigma}$) is the electron creation (annihilation) operators at the $i$-th site with spin state $\sigma= \ket{\uparrow}, \ket{\downarrow}$; $\taus[i=0,x,y,z]$ are the Pauli matrices which acts only on  the spin states; $t$ is the electron hopping energy; $\alpha$ is the RSOC strength; $\mu$ is the chemical potential; $B_{z}$ is the externally applied magnetic field perpendicular to the NW (along $z$-axis); $\Delta$ is the extended $s$-wave SC order parameter; $U$ is the on-site Coulomb repulsion energy; $n_{i \sigma}=c_{i \sigma}^{\dagger}c_{i \sigma}$ is the electron number operator at $i$-th site.

In case $U$ is a dominant energy scale, the Hubbard $U$ term enforces strong correlation: the low-energy sector of the underlying on-site Hilbert space is essentially modified. Namely, in the limit $U \to \infty$ the double electron occupancy is strictly prohibited so that the original four dimensional onsite Hilbert space spanned by the states $\left( \ket{\uparrow},\ket{\downarrow},\ket{\uparrow \downarrow}, \ket{0}\right)$, reduces to a three dimensional Hilbert space spanned by the states $\left( \ket{\uparrow}, \ket{\downarrow} \ket{0} \right)$. Such a restriction results in entirely new low-energy excitations. To account for this new physics one uses projected electron operators
\begin{equation}
  \label{eq:hubbard-opr}
  \begin{aligned}
    &\mathcal{P}_ic_{i\sigma} \mathcal{P}_i = (1-n_{i \sigma'})c_{i\sigma}=:X_{i}^{0\sigma} ,\\
    & \mathcal{P}_ic_{i\sigma}^{\dagger}\mathcal{P}_i =c_{i\sigma}^{\dagger} (1-n_{i \sigma'})=:X_{i}^{\sigma 0}.
  \end{aligned}
\end{equation}
Here, $X_i^{pq}=\ket{p}\bra{q}$ with $p,q \equiv \left(\uparrow,\downarrow,0 \right)$, are Hubbard operators represented now by a $3\times 3$ matrices. $\mathcal{P}_i= 1-n_{i\uparrow}n_{i\downarrow}$ is the so-called Gutzwiller projection operator, it satisfies the condition $\mathcal{P}_i^2=\mathcal{P}_i$. In the strong correlation limit the NW Hamiltonian, Eq. \ref{eq:basic-ham}, reduces to
\begin{equation}
  \label{eq:strong-ham}
  \begin{aligned}
\mathcal{P} H \mathcal{P}=&-X_i^{\sigma 0}\left(t \tau_{0}^{\sigma\sigma'} + \alpha \tau_{y}^{\sigma\sigma'} \right) X_{i+1}^{0\sigma'}
-\imath X^{\sigma 0}_{i}\Delta \tau_{y}^{\sigma\sigma'}X^{\sigma' 0}_{i+1}\\
&\quad +\text{H.c.} + X_{i}^{\sigma 0} \left( \mu \tau_{0}^{\sigma\sigma'} + B_{z}
\tau_{z}^{\sigma\sigma'} \right) X_{i+1}^{0\sigma'}.
\end{aligned}
\end{equation}
The Gutzwiller operator on L.H.S is defined as $\mathcal{P}=\prod_i\mathcal{P}_i$. It should be noted that, as $X_i^{\sigma 0}X_i^{\sigma' 0} \equiv 0$, the NDO constraint is resolved explicitly.

The important point is that a full set of Hubbard operators $X^{pq}$ (we dropped index $i$ for convenience) is closed into the $su(2|1)$ superalgebra in the fundamental representation \cite{wiegmann-1988-super-stron} (we recommend Ref. \cite{ferraz_2011_EffectiveAction_NuclearPhysicsB} for an in-depth introduction to the $su(2|1)$ path integral method applied here). This implies that a powerful method of the coherent state path integral quantization \cite{zhang-1990-coher-states} can be applied to treat a strongly correlated electron system \footnote{The applied method heavily relies on the coherent state path integral approach. A good introduction to the subject can be found in the lectures by L. D. Faddeev \cite[Pg. 80-118 of ][]{faddeev-1995-40-years}, book by L. S. Schulman \cite[Chapter 27 of ][]{schulman-2005-techniques-application}, collection of articles and specifically introduction part of Ref. \cite[Chapter I of ][]{klauder-1985-cohrent-states}, works by A. M. Perelomov \cite{perelomov-1977-gener-coher,perelomov-1986-gener-coher}. Any reference on this subject can not be complete without the mention of the works of F. A. Berezin \cite{berezin_1987_IntroductionSuperanalysis_,berezin-1980-feynm-path} on superalgebra.}. Namely, given a Hamiltonian in terms of the Hubbard operators the corresponding imaginary time phase space action takes on the form,
\begin{equation}
  \label{eq:action-phase-space}
  \begin{aligned}
    \mathcal{A}=-\int\limits_0^{\beta} \left\langle z, \xi \left| \frac{\partial}{\partial t} + H \right|z,\xi \right\rangle dt,
  \end{aligned}
\end{equation}
where
\begin{equation}
  \label{eq:cs-symbol-su-1}
  \begin{aligned}
    \ket{z, \xi}& \equiv \left[\frac{\exp \left( z \xopr{\downarrow}{\uparrow} + \xi \xopr {0}{ \uparrow}\right)}{\sqrt{ 1 + \bar{z}z + \bar{\xi} \xi}}  \right] \ket{\uparrow} = \frac{ \ket{\uparrow}
                  + z \ket{\downarrow} + \xi \ket{0}}{\sqrt{ 1 + \bar{z}z + \bar{\xi} \xi}}.
  \end{aligned}
\end{equation}
being the normalized coherent state associated with the lowest irreducible representation of the $su(2|1)$ superalgebra spanned by Hubbard operators \cite{kochetov_1995_SUCoherent_JournalOfMathematicalPhysics,ferraz_2011_EffectiveAction_NuclearPhysicsB}. The partition function takes a form of the $su(2|1)$ coherent state path integral:
\begin{equation}
  \label{eq:part-function}
Z=\int D \: \mu_{su(2|1)} (z,\xi ) \exp \mathcal{A}.
\end{equation}
Here $D \: \mu_{su(2|1)} (z,\xi )$ is the measure. The dynamical field $z$ is a complex time-dependent variable and $\xi$ is an odd complex time-dependent Grassmann variable that characterize the inhomogeneous (proper) coordinates of a point on a supersphere \cite[See chapter 2 of][]{fayet-1977-super}, i.e.
$$(z,\xi) \in S^{2|2}\simeq CP^{1|1}=SU(2|1)/U(1|1).$$
Here $CP^{1|1}$ stands for a complex projective superspace with a complex dimension $(1,1)$. It can be thought of as the minimal superextension of an ordinary projective space $CP^1$ -- a phase space of the $su(2)$ spin. The odd Grassmann parameter appears in Eq. (\ref{eq:cs-symbol-su-1}) due to the fact that $X^{\downarrow 0}$ is a fermionic operator in contrast with the bosonic operator $X^{\downarrow\uparrow}$. The product $\xi X^{0\uparrow}$ represents therefore a bosonic quantity as required. The $CP^{1|1}$ manifold serves as a classical phase space for the Hubbard operators.

The effective Hamiltonian (\ref{eq:strong-ham}) in terms of the $su(2|1)$ coherent state symbols reads (see App. \ref{sec:su21} for derivation):
\begin{equation}
  \label{eq:effectiv-ham-trans}
  \begin{aligned}
    H_{\text{eff}}(z, \xi) =&- \sum\limits_{i} \xi_{i+1} \bar{\xi}_{i} \left(t \: a_{i} + \imath \alpha \: \alpha_{i} \right) - \Delta\sum\limits_{i}\bar{\xi}_{i+1}\bar{\xi}_{i}\: \Delta_{i}\\
                            & \quad + \text{H.c.} + \sum\limits_{i} \xi_{i} \bar{\xi}_{i} \left(\mu + B_{z}\beta_{i}\right),
  \end{aligned}
\end{equation}
where,
\small
\begin{equation}
  \label{eq:def-aij}
  \begin{aligned}
    &a_{i} \equiv \frac{1 + \bar{z}_{i+1} z_{i}}{\sqrt{\left( 1 + \left| z_{i+1} \right|^{2}\right)\left( 1 + \left| z_{i} \right|^{2}\right)}},
      \alpha_{i} \equiv \frac{\bar{z}_{i+1} - z_{i}}{\sqrt{\left( 1 + \left| z_{i+1} \right|^{2}\right) \left( 1 + \left| z_{i} \right|^{2}\right)}},\\
    &\Delta_{i} \equiv \frac{z_{i} - z_{i+1}}{\sqrt{\left( 1 + \left| z_{i} \right|^{2}\right) \left( 1 + \left| z_{i+1} \right|^{2}\right)}}, \beta_{i} \equiv \frac{1 - \left| z_{i}^{2} \right|}{1 + \left| z_{i}^{2} \right|}.
  \end{aligned}
\end{equation}
\normalsize
Parameters $a_{i}$, $\alpha_{i}$, $\Delta_{i}$ and $\beta_{i}$ are functions of $z_{i}$, which renders these coefficients spatially dependent. The fractionalization of the spin (spinon, $z_{i}$) and charge (holon, $\xi_{i}$) degrees of freedom in Eq. (\ref{eq:effectiv-ham-trans}) allows them to be treated on the same footing. Such a fractionalization is believed to be a hallmark of strong correlation \cite{senthil-2000}.

To proceed we suggest that, in the low-energy sector, the direct charge-spin mixing is not very important. This is applicable provided the elementary charge and spin excitations are separated by a large energy gap so that the spin and  charge energy scales are essentially different from each other. Which is believed to be the case for strongly correlated lattice electrons at small doping, $ \left \langle \bar\xi_i\xi_i \right \rangle \ll 1 $ (e.g. in the lightly doped $t-J$ model of high-$T_c$ superconductivity). The relevant field transformations are then those generated by the even subgroup $SU(2)\times U(1)$ so that the $(1-\bar{\xi}\xi)$ factor in the measure can be dropped. This implies that in the low energy regime the fields $z_i(t)$ and $\xi_i(t)$ which appear in Eq. (\ref{eq:strong-ham}) describe well-defined $su(2)$ spin and $u(1)$ spinless fermion collective excitations respectively.

\section{spin structures}

As the spin ($z$) and charge ($\xi$) degrees of freedom in the effective Hamiltonian, Eq. (\ref{eq:effectiv-ham-trans}), have been separated, therefore number of physical problems can be investigated by representing the spin structures (viz. ferromagnetic, antiferromagnetic, spiral, conical etc.) through spinon fields.

The coherent state $su(2)$ generators of the three spin components in terms of $z_{i}$ are (see App. \ref{sec:su21}):
\begin{equation}
  \label{eq:spin-cov-symb}
  \begin{aligned}
    S_{i}^{+} = \frac{z_{i}}{1+\left| z_{i} \right|^{2}},\quad
    S_{i}^{-} = \frac{\bar{z}_{i}}{1+\left| z_{i} \right|^{2}},\quad
    S_{i}^{z} = \frac{1}{2} \left( \frac{1- \left| z_{i} \right|^{2}}{1+\left| z_{i} \right|^{2}} \right).
  \end{aligned}
\end{equation}
Here, $S_{i}^{x}=S_{i}^{+}+S_{i}^{-}$ and $S_{i}^{y}=-\imath \left(S_{i}^{+} -S_{i}^{i} \right)$. As $z_{i}$ is a $c$-number, in Cartesian form its real and imaginary parts are:
\begin{equation}
  \label{eq:z-real-imag-cart}
  \begin{aligned}
    \Re z_{i} = S_{i}^{x} \left( 1 + \frac{1 - 2 S_{i}^{z}}{1 + 2 S_{i}^{z}} \right), \quad \Im z_{i} = S_{i}^{y} \left( 1 + \frac{1 - 2 S_{i}^{z}}{1 + 2 S_{i}^{z}} \right).
  \end{aligned}
\end{equation}
In polar form $z_{i}$ can be represented as
\begin{equation}
  \label{eq:z-real-imag-polar}
  \begin{aligned}
    z_{i}=\left( 1 + \frac{1 - 2 S_{i}^{z}}{1 + 2 S_{i}^{z}} \right) \sqrt{\left(S_{i}^{x}\right)^{2} + \left(S_{i}^{y} \right)^{2}} \exp \left[\imath \arctan \left( S_{i}^{y}/S_{i}^{x} \right) \right].
  \end{aligned}
\end{equation}
One can use Eqs. (\ref{eq:spin-cov-symb}), (\ref{eq:z-real-imag-cart}), (\ref{eq:z-real-imag-polar}) to recast the Hamiltonian, Eq. (\ref{eq:effectiv-ham-trans}), in a more familiar form containing three spin components.

For a ferromagnetic (FM) configuration all the spins are aligned along the $z$-axis:
\begin{equation}
  \label{eq:spin-config-spiral-fm}
  \vec{S}_{i} = \left( S_{i}^{x}, S_{i}^{y}, S_{i}^{z} \right) = \frac{1}{2}\left(0,0, 1  \right).
\end{equation}
Substituting Eq. (\ref{eq:spin-config-spiral-fm}) in Eq. (\ref{eq:z-real-imag-cart}) will give $z_{i}=0$. It means in Eq. (\ref{eq:def-aij}) only $a_{i}$ and $\beta_{i}$ are non-zero. The effective Hamiltonian, Eq. (\ref{eq:effectiv-ham-trans}), reduces to usual 1D NW without SC and RSOC. Therefore FM configuration is not of our interest. The antiferromagnetic spinon configuration will also give the same result \footnote{For AFM case $\vec{S}_{i}= \left( S_{i}^{x}, S_{i}^{y}, S_{i}^{z} \right)= (0,0, \pm 1/2)$. The spin lattice is divided into two FM sub-lattices with opposite spin orientation. Therefore the individual effective Hamiltonian of the two sub-lattices will give result in no RSOC and SC as in FM case. Another method to reach the same conclusion is to use $S_{i}^{x}=S_{i}^{y}=0$, which will result in $z_{i}=0$ from Eq. \eqref{eq:z-real-imag-polar}.}. However, spiral spinon field is different. The spatially dependent spinon configuration having only \emph{x-y} plane projection reads
\begin{equation}
  \label{eq:spin-config-spiral-xy}
  \vec{S}_{i} = \left( S_{i}^{x}, S_{i}^{y}, S_{i}^{z} \right) = \frac{1}{2}\left(\cos \theta_{i}, \sin \theta_{i}, 0  \right).
\end{equation}
Here the $S_{i}^{z}$ component of the spinon is zero and $\theta_{i}$ is spatially dependent. The corresponding $c$-number obtained from Eq. (\ref{eq:z-real-imag-polar}) is $z_{i}=\mathrm{e}^{\imath \theta_{i}}$. Substituting this $z_{i}$ in Eq. (\ref{eq:def-aij}) we find:
\begin{equation}
  \label{eq:spiral-spin-aij}
  \begin{aligned}
    &a_{i} =  \mathrm{e}^{-\imath \left(\theta_{i+1}-\theta_{i}\right)/2}  \cos \frac{\left(\theta_{i+1}-\theta_{i} \right)}{2}, \quad \beta_{i} =0,\\
    &\alpha_{i} = -\imath \mathrm{e}^{-\imath \left(\theta_{i+1}-\theta_{i}\right)/2}  \sin \frac{\left(\theta_{i+1}+\theta_{i} \right)}{2},\\
    &\Delta_{i} =  -\imath \mathrm{e}^{\imath \left(\theta_{i+1}-\theta_{i}\right)/2}  \sin \frac{\left(\theta_{i+1}-\theta_{i} \right)}{2}\mathrm{e}^{\imath \theta_{i}}.
  \end{aligned}
\end{equation}
For a spiral spinon field the difference between the $\theta_{i}$ and $\theta_{i+1}$ is constant, i.e. $\theta_{i+1}-\theta_{i}=\theta$; in other words $\theta_{i}= i \times \theta$. Using this fact in Eq. (\ref{eq:spiral-spin-aij}), substituting the resulting expressions in Eq. (\ref{eq:effectiv-ham-trans}), and making the gauge transformation $\xi_{i} \to \xi_{i}\mathrm{e}^{\imath (\theta_{i}/2-\pi/4)}$, we get the Hamiltonian with a spiral spinon field:
\begin{equation}
  \label{eq:effectiv-ham-spiral-spin}
  \begin{aligned}
    H_{\text{eff}}=-\sum\limits_{i} \xi_{i+1} \bar{\xi}_{i} \: \tilde{t}_{i} - \Delta \sin \frac{\theta}{2}\sum\limits_{i}\bar{\xi}_{i+1}\bar{\xi}_{i} + \text{H.c.} + \mu\sum\limits_{i} \xi_{i}\bar{\xi}_{i},
  \end{aligned}
\end{equation}
where,
\begin{equation*}
  \tilde{t}_{i} = t \cos \frac{\theta}{2} + \alpha \sin \left( \theta_{i} + \frac{\theta}{2} \right).
\end{equation*}
Comparing Eqs. (\ref{eq:effectiv-ham-trans}) and (\ref{eq:effectiv-ham-spiral-spin}) we notice that the term containing the magnetic field is absent, hence, we find an analogous Hamiltonian, Eq. (\ref{eq:effectiv-ham-spiral-spin}), even if in the original Hamiltonian, Eq. (\ref{eq:basic-ham}), there is an applied magnetic field. We also note that in Eq. (\ref{eq:effectiv-ham-spiral-spin}) the electron hopping ($\tilde{t}_{i}$) has a sinusoidal periodic character. From now on we explicitly investigate the Hamiltonian with the spiral spin structure, as it has been observed in number of vdW materials and NW (see Sec. \ref{sec:heter-exper-real} and \ref{sec:gener-spir-spin}). The case for the conical spin structure is discussed in Sec. \ref{sec:conic-spin-struct}.

\section{Topological invariants: $\mathbb{Z}_{2}$ and $\mathbb{Z}$ }

Eq. (\ref{eq:effectiv-ham-spiral-spin}) is analogous to the 1D Kitaev chain \cite{kitaev_2001_UnpairedMajorana_Phys-Usp} with a modulated electron hopping. Hence, one can expect that the system has rich topological properties. Further on $\xi_{i}$ is treated as spinless electrons, as the spin degrees of freedom is taken into account in spinon fields $z_{i}$. If $\tilde{t}_{i}$ has period of $N= 2\pi/\theta$ sites (assuming $\theta=2\pi/N$), the first BZ of Eq. (\ref{eq:effectiv-ham-spiral-spin}) is folded $N$ times and it has the boundary $\left( -\pi/N, \pi/N \right)$. The Hamiltonian in momentum space with periodic $\tilde{t}_{i}$ and periodic boundary condition is
\begin{equation}
  \label{eq:ham-mom}
  \begin{aligned}
    H = \sum\limits_{k} \Psi_{k}^{\dagger} H_{k} \Psi_{k};
        \quad H_{k} =
            \begin{bmatrix}
              M_{k} &\Delta_{k}\\
              -\Delta_{k}^{\dagger} &-M_{k}^{\dagger}
            \end{bmatrix};\\
    \Psi_{k}^{\dagger}= \left( c_{1,k}^{\dagger},\dots, c_{N,k}^{\dagger},c_{1,-k},c_{N,-k}\right).
  \end{aligned}
\end{equation}
Here $H_{k}$ is a $2N \times 2N$ matrix. $M_{k}$ and $\Delta_{k}$ are $N \times N$ matrices; for $i \in \left( 1,N-1 \right)$ the elements are $M_{k}^{i,i+1}=M_{k}^{i+1,i}=-\tilde{t}_{i}$, $M_{k}^{i,i}=\mu/2$, $\Delta_{k}^{i,i+1}=-\Delta_{k}^{i+1,i}=\Delta \sin \left( \theta/2 \right)$; for $i=N$ the elements are $M_{k}^{N,N}=\mu/2$, $M_{k}^{N,1}=-\tilde{t}_{N} \mathrm{e}^{-\imath N k}$, $M_{k}^{1,N}=-\tilde{t}_{N} \mathrm{e}^{+\imath N k}$, $\Delta_{k}^{N,1}=\Delta \sin \left( \theta/2 \right) \mathrm{e}^{-\imath N k}$, $\Delta_{k}^{1,N}=-\Delta \sin \left( \theta/2 \right) \mathrm{e}^{+\imath N k}$. The $k$ lies in the first BZ with $k \in \left( -\pi/N, \pi/N \right)$.

The time reversal symmetry ($\Theta=K$, where $K$ is complex conjugation), the particle hole symmetry ($\Xi= \tau_{x}K$) and chiral symmetry ($\Pi=\tau_{x}$) are all conserved in $H_{k}$. Hence, using the unitary transformation with $U=\mathrm{e}^{i \tau_{y} \pi/4}$ we can represent $H_{k}$ in the off diagonal form \cite{ryu-2010-topol-insul-super}:
\begin{equation}
  \label{eq:U-trans}
  UH_{k}U^{\dagger} =
  \begin{bmatrix}
    0 &A_{k}\\
    A_{k}^{\dagger} &0
  \end{bmatrix},
  \: A_{k} = M_{k} + \Delta_{k},
  \: U = \frac{1}{\sqrt{2}}
  \begin{bmatrix}
    I &-I\\
    I & I
  \end{bmatrix}.
\end{equation}
The system is a BDI TSC \cite{ryu-2010-topol-insul-super}. The number of Majorana zero modes (MZM) present at the end of the wire depends on the explicit value of $A_{k}$, given by the winding number ($W$) \cite{tewari-2012-topol-invar}:
\begin{equation}
  \label{eq:wind-num}
  W= \frac{-i}{\pi} \int\limits_{k=0}^{k=\pi/N} d z_{k}/z_{k}; \quad z_{k} = \text{Det} \left( A_{k} \right)/\left| \text{Det} \left( A_{k} \right) \right|.
\end{equation}
Here $W$ essentially counts how many times $\text{Det}\left( A_{k} \right)$ crosses the imaginary axis. We can also count the $\mathbb{Z}_{2}$ Pfaffian invariant ($Q$) of the system which is just the parity of the $\mathbb{Z}$ invariant ($W$) \cite{tewari-2012-topol-invar}:
\begin{equation}
  \label{eq:z-2-invariant}
  Q = \text{sgn} \left[ \frac{\text{Det} \left\{ A_{k=\pi/N} \right\}}{\text{Det} \left\{ A_{k=0} \right\}} \right] = (-1)^{W}.
\end{equation}
$Q=1$ ($Q=-1$) is topologically trivial (non-trivial) phase. Therefore, a topological non-trivial phase occurs when $W$ is odd. Physically $W$ characterizes the number of MZM present. Hence, a topological non-trivial phase occurs only when odd number of MZM is present. For even number of MZM, the MF can be combined to create usual fermion modes.

\section{Re-entrant topological phase}
\label{sec:phase-diagram}
\begin{figure}[tbh]
  \centering
  \includegraphics[width=0.48\textwidth]{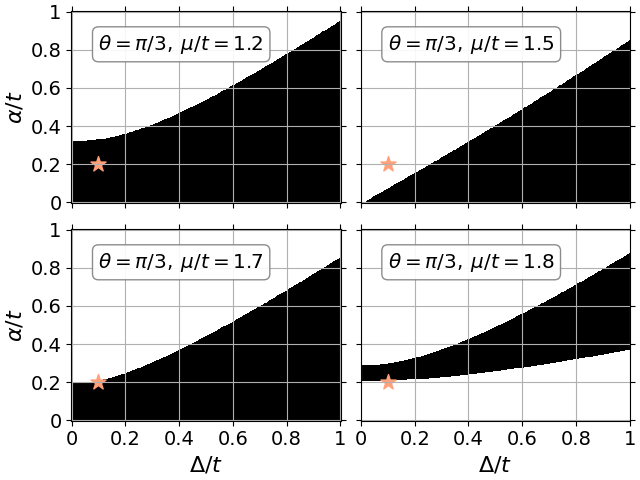}
  \caption{Color online: Dependence of $\mathbb{Z}_{2}$ topological invariant on SC gap ($\Delta$) and RSOC ($\alpha$) at chemical potential ($\mu/t=1.2,1.5,1.7,1.8$) for  spiral spinon texture with phase modulation angle $\theta=\pi/3$. The topologically trivial (non-trivial) phase is shown in white (black). The position $\alpha/t=0.2$ and $\Delta/t=0.1$ is marked by red star to point out the re-entrant topological transition. In Fig. \ref{fig:seq-phase-diag} the phase diagram for extended parameter region is shown.}
  \label{fig:seq-phase-diag-new}
\end{figure}

The topological invariant $Q$ is related to $A_{k}$, which is a function of parameters $\mu$, $\theta$, $\Delta$ and $\alpha$. Therefore an interesting topological phase diagram can be found by tuning accordingly these parameters. In Fig. \ref{fig:seq-phase-diag-new} we plot the dependence of $Q$ on these parameters (see also Fig. \ref{fig:seq-phase-diag} for extended parameter range). Specifically, we plot the topological invariant $Q$ on the $\Delta$--$\alpha$ parameter space for $\mu/t=1.2,1.5,1.7,1.8$ and $\theta=\pi/3$. It can be observed that with increase in $\mu$ the topologically trivial phase (white) starts to seep in from the upper left corner (higher $\alpha$ and lower $\Delta$); it is clearly visible in Fig. \ref{fig:seq-phase-diag} where phase space for extended parameter range is shown. This can be understood by keeping in mind the energy spectrum \footnote{For an archetypal Kitaev chain the energy spectrum has a cosine dependence \cite{alicea_2012_NewDirections_RepProgPhys}. The topological boundary corresponds to $\left| \mu \right| \leq 2t$. Physically it corresponds to the maximum ($2t$) and minimum ($-2t$) of the energy band.} of the archetypal Kitaev chain \cite{kitaev_2001_UnpairedMajorana_Phys-Usp,alicea_2012_NewDirections_RepProgPhys} and by observing the momentum space energy spectrum of the Hamiltonian, Eq. (\ref{eq:ham-mom}). Although Eq. (\ref{eq:effectiv-ham-spiral-spin}) is analogous to the Hamiltonian of the archetypal Kitaev chain in real space, but in momentum space the energy spectrum differs from each other due to presence of $\theta$. Effect of inclusion of $\theta$ in the Hamiltonian are of two folds. First, with the increase in $\theta$ the energy dispersion becomes flatter (hopping is proportional to $t \cos (\theta/2)$, hence the width of band decreases), therefore, the electrons become more localized. Second, when $\alpha \neq 0$, depending on periodicity of $\tilde{t}_{i}$ (periodicity depends on $\theta$) corresponding number of band gaps appear in the energy spectrum; when $\alpha=0$ only a single band is present as $\tilde{t}_{i}$ loses its periodicity.

\begin{figure}[tbh]
  \centering
  \includegraphics[width=0.48\textwidth]{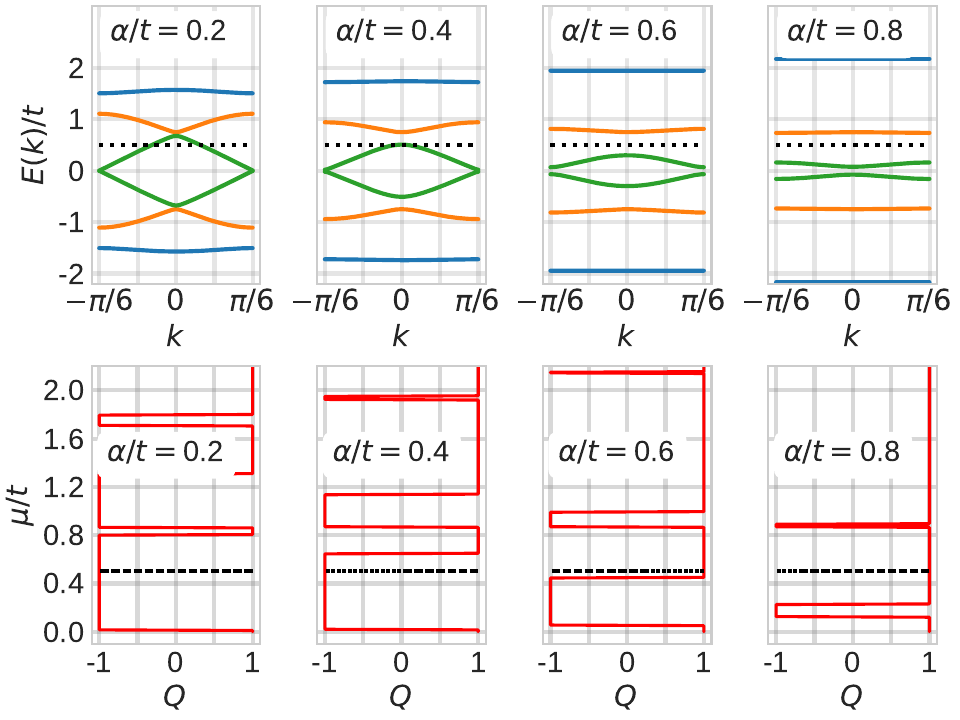}
  \caption{(Upper plots) The energy spectrum in a reduced BZ for $\theta=\pi/3$ and $\Delta/t=0.01$. The BZ boundary is at $\pi/6$ as the periodicity of $\tilde{t}_{i}$ is sixfold. Depending on the value of $\alpha$, the $\mu$ may be present inside or outside of the band (in the bandgap), e.g. for $\alpha/t =0.2$ the $\mu/t=0.5$ (dotted, black) level is inside the band, however, it is in the bandgap for $\alpha/t=0.8$. (Lower plots) The $x$-axis represent $\mathbb{Z}_{2}$ topological invariant $Q$, and $y$-axis the chemical potential $\mu/t$ for $\theta=\pi/3$, $\Delta/t=0.01$, and $\alpha/t=0.2,0.4,0.6,0.8$. The $\mu/t=0.5$ (dotted, black) shows the presence (for $Q=-1$) of topological phase at $\alpha/t=0.2,0.4$, and the absence (for $Q=1$) of the topological phase at $\alpha/t=0.6,0.8$.}
  \label{fig:ener-spect-alpha-dep}
\end{figure}

In Fig. \ref{fig:ener-spect-alpha-dep} we plot the energy spectrum for $\theta=\pi/3$, $\Delta/t=0.01$ and $\alpha/t=0.2,0.4,0.6,0.8$ for the Hamiltonian, Eq. \eqref{eq:ham-mom}. The small value of $\Delta$ is taken for the convenience of the readers. The small $\Delta$ allows us to concentrate only on the left margin of the subplots for $\theta=\pi/3$ in Fig. \ref{fig:seq-phase-diag}. Readers can compare the corresponding topological phase diagram in Fig. \ref{fig:seq-phase-diag} (for $\theta=\pi/3$) with Fig. \ref{fig:ener-spect-alpha-dep}; for example, by going upwards on the left margin (increase in $\alpha$) of a subplot in Fig. \ref{fig:seq-phase-diag}, or going from left to right (increase in $\mu/t$) in the row for $\theta=\pi/3$. It should be noted that due to small $\Delta$ the superconducting energy gap at chemical potential in Fig. \ref{fig:ener-spect-alpha-dep} is not clear, however, it is present. We see that six energy bands are present for each $\alpha$; it was expected as periodicity of $\tilde{t}_{i}$ in this case is sixfold. Besides, we also observe that with the increase in $\alpha$ the energy gap increases, and bands become flatter. Now the reason for the onset of the trivial phase from the values $\alpha \approx 1, \Delta \approx 0$ (upper left corner of each row) in Fig. \ref{fig:seq-phase-diag} is clear. At high $\alpha$ the bands are flatter compared to the ones with low $\alpha$. Therefore, when for higher $\alpha$ we gradually increase $\mu$ the maximum of the band is reached first, compared to the lower $\alpha$ case. For example in Fig. \ref{fig:ener-spect-alpha-dep} when $\alpha/t=0.8$ the $\mu/t=0.5$ level already lies inside the band gap between the first and the second bands (the bands are counted from $\mu/t=0$), whereas, for $\alpha/t=0.2$ it lies inside the first band. Therefore in the Fig. \ref{fig:seq-phase-diag} for $\theta=\pi/3$, $\Delta/t \approx 0$ and $\mu/t=0.5$ the system is already in trivial topological phase for $\alpha/t=0.8$, although, for $\alpha/t=0.2$ the topological non-trivial phase is still present in the system.

Another interesting feature we observe in Fig. \ref{fig:seq-phase-diag-new} (see also Fig. \ref{fig:seq-phase-diag}), is the multiple transition from topologically trivial to non-trivial phase (and vice-versa) as the parameters are varied (we call it ``\emph{re-entrant}'' topological phase transition). For example the system for $\theta=\pi/3$, $\alpha/t=0.2$, $\Delta/t=0.1$ (red star) goes through phase transitions three times as $\mu$ increases: (i) at $\mu/t=1.5$ from non-trivial to trivial phase, (ii) at $\mu/t=1.7$ from trivial to non-trivial phase, (iii) at $\mu/t=1.8$ from non-trivial to trivial. This can be qualitatively understood from the energy spectrum diagram of the system. As $\mu$ is increased the system goes through all the bands. The transition from non-trivial (trivial) to trivial (non-trivial) phase occurs when $\mu$ goes from inside the band (band gap) into the band gap (band). In the band gap there is no electronic states to generate MFs, hence, the phase is trivial. This is analogous to the behavior of the archetypal Kitaev chain \cite{kitaev_2001_UnpairedMajorana_Phys-Usp}. In Kitaev chain due to absence of the periodicity only single band is present. When the $\mu$ lies inside the band $-2<\mu/t<2$ then non-trivial phase is present. However, when $\mu$ lies outside this range the phase is trivial. In our case due to the periodic nature of the Hamiltonian (when $\alpha >0 $) the BZ gets folded ($N$ times if periodicity is $N$), which results in the emergence of additional band gap at the boundary of the folded BZ. If the $\mu$ lies inside these band gap then the phase is trivial; if the $\mu$ lies inside the band then phase is non-trivial. In fact, not only $\mu$, but also, by varying $\Delta$ (blue down triangles in first row of Fig. \ref{fig:seq-phase-diag}, $\theta=\pi/6$, $\mu/t=1.2$) and $\alpha$ (green up triangles in first row of Fig. \ref{fig:seq-phase-diag}, $\theta=\pi/6$, $\mu/t=1.8$) the re-entrant topological phase transition can be achieved.

\section{Numerical simulation}
\label{sec:numerical-simulation}
\begin{figure}[tbh]
  \centering
  \includegraphics[width=0.48\textwidth]{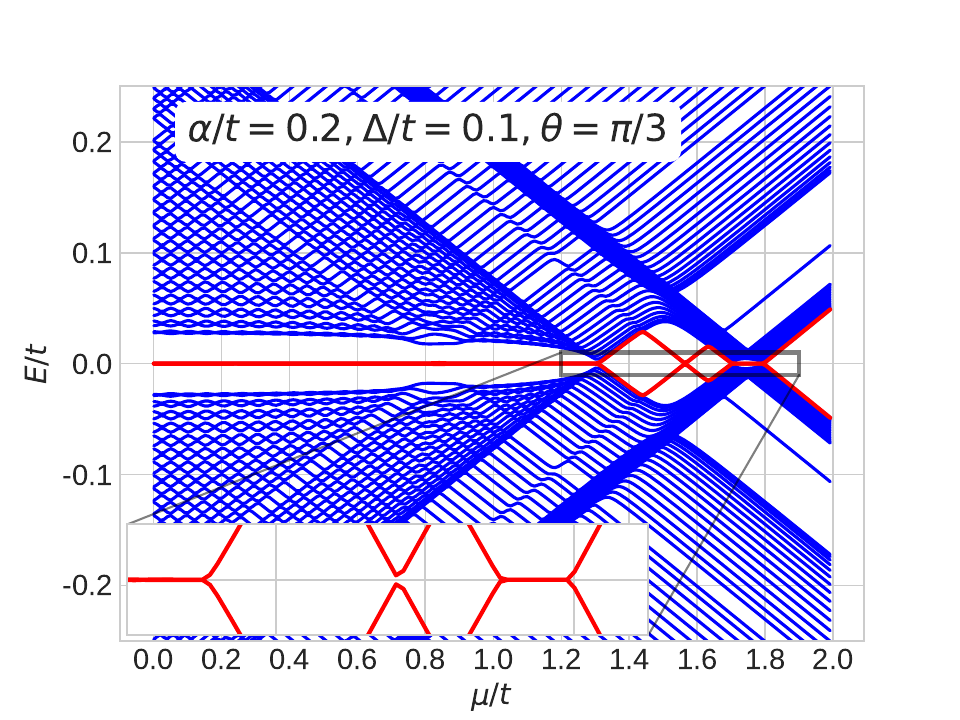}
  \caption{The energy spectrum found by diagonalizing Eq. (\ref{eq:effectiv-ham-spiral-spin}) in the Majorana basis under an open boundary condition for a chain of length $L=100$., $\theta=\pi/3$, $\Delta/t=0.1$ and $\alpha/t=0.2$; these values of the parameters are same as the marked (red, star) position in Fig. \ref{fig:seq-phase-diag} for $\pi/3$. (inset) Zoomed portion of energy spectrum around zero energy level from $\mu/t=1.2$ to $1.9$. It is shows that at $\mu/t \approx 1.5$ the MZM really no longer appears.}
  \label{fig:kitaev-chain-dep-on-mu}
\end{figure}

To make sure that topological phase really occurs in these systems we perform numerical simulations. The signature of a topological phase is the occurrence of two sets of spatially separated odd numbers of MF at two ends of the wire under an open boundary condition; they are represented by MZM states in energy spectrum \cite{alicea_2012_NewDirections_RepProgPhys}. In Fig. \ref{fig:kitaev-chain-dep-on-mu} we numerically diagonalize the Hamiltonian, Eq. (\ref{eq:effectiv-ham-spiral-spin}), under open boundary condition in the Majorana basis. The parameters are $\alpha/t=0.2$ and $\Delta/t=0.1$, is same as the marked points in Fig. \ref{fig:seq-phase-diag} with $\theta=\pi/3$ (red star)[in Fig. \ref{fig:seq-phase-diag-new} we showed part of Fig. \ref{fig:seq-phase-diag}]. We observe that going from right to left in Fig. \ref{fig:seq-phase-diag} at these parameters the system undergoes through three phase transitions. In the numerical simulation in Fig. \ref{fig:kitaev-chain-dep-on-mu} we also observe these three phase transitions (destruction and creation of zero modes).

\begin{figure}[tbh]
  \centering
  \includegraphics[width=0.48\textwidth]{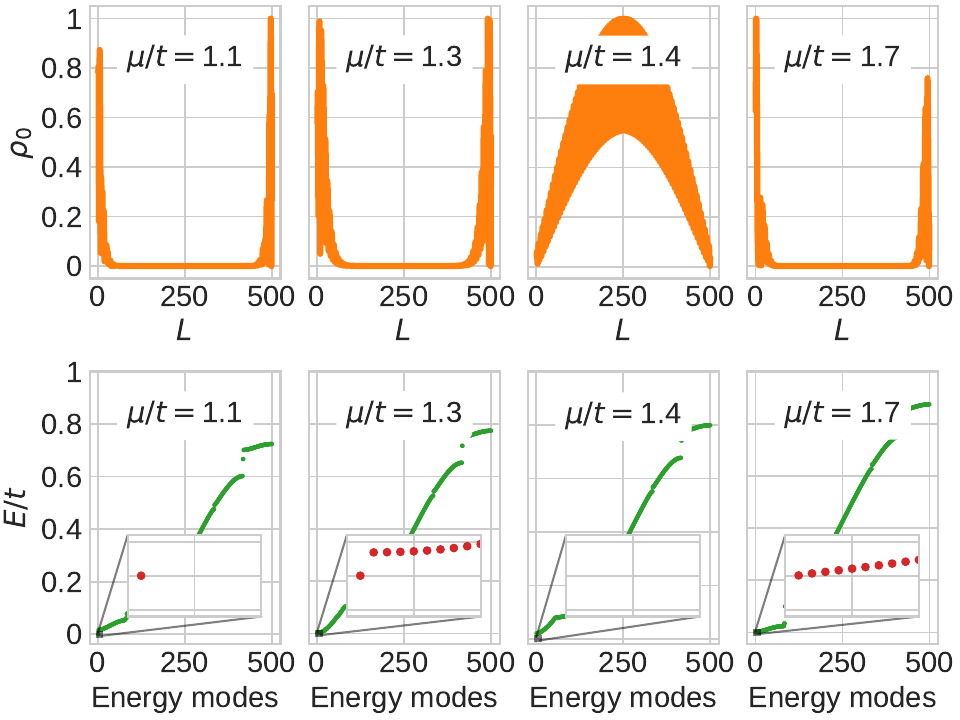}
  \caption{(Upper plots) Normalized LDOS ($\rho_{0}$) of the lowest energy modes along the wire. $\rho_{0}$ is found by diagonalizing (in a Nambu basis) the Hamiltonian, Eq. (\ref{eq:effectiv-ham-spiral-spin}), for a wire of length $L=500$, $\theta=\pi/3$, $\Delta/t=0.2$, $\alpha/t=0.1$. The two peaks at the two ends of the wire for $\mu/t=1.1,1.3,1.7$ shows the presence of spatially separated Majorana fermions. When $\mu/t=1.4$ the Majorana fermions are not localized at the end of the wire, but they can instead be found in the bulk. (Lower plots) The positive energy modes calculated by diagonalizing (in the Majorana basis) the Hamiltonian for the same parameter values. (inset) Shows the presence of zero energy modes for $\mu/t=1.1,1.3,1.7$, and absence of the zero energy modes at $\mu/t=1.4$.}
  \label{fig:majorana-fm-ener-modes}
\end{figure}

To determine the positions of the zero modes, i.e. whether they really are localized at the end of the wire or not, we numerically diagonalized the real space Hamiltonian (in Nambu basis) of a chain of length $L=500$ and find the localized density of states (LDOS) of the zeroth mode energy ($\rho_{0}$) \cite{sacramento-2007-magnet-impur-super} for $\theta=\pi/3$, $\alpha/t=0.1$, $\Delta/t=0.2$ and $\mu/t=1.1,1.3,1.4,1.7$; in Fig. \ref{fig:seq-phase-diag} these parameters are also marked (red star). In Fig. \ref{fig:majorana-fm-ener-modes} the presence of two peaks at the two ends of the wire is due to localization of the MFs at the two end of the wire for $\mu/t=1.1,1.3,1.7$. However, for $\mu/t=1.4$ the MFs are not localized at the end of the wire and can be found in the bulk of the wire; hence, they are not topologically protected. The same behavior is shown in the calculated phase diagram in Fig. \ref{fig:seq-phase-diag}. To make sure, that the end states are really MZM, in Fig. \ref{fig:majorana-fm-ener-modes} (lower plots) we also show the energy modes of the same wire with same parameter configurations. We can clearly observe that the zero modes are present for $\mu/t=1.1,1.3,1.7$ and absent in $\mu/t=1.4$.

\section{Dynamic Ground states}
\label{sec:dynam-ground-stat}
\begin{figure}[tbh]
  \centering
  \includegraphics[width=0.48\textwidth]{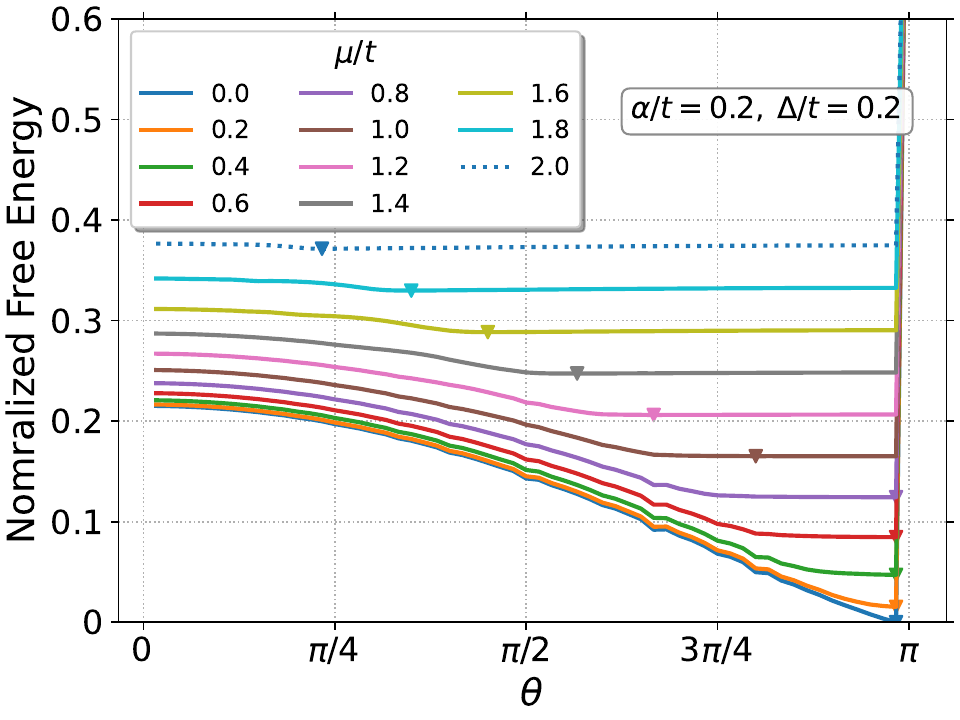}
  \caption{Dependence of free energy of the Hamiltonian, Eq. (\ref{eq:effectiv-ham-spiral-spin}), on chemical pontential $\mu/t$ and spiral modulation angle $\theta$. The minimum energy $\theta$ is marked by the down triangles.}
  \label{fig:min-ener-plt}
\end{figure}
In the investigated Hamiltonian, Eq. (\ref{eq:effectiv-ham-spiral-spin}), $\alpha$, $\Delta$, $\mu$ and $\theta$ define the existing topological properties. Among these parameters $\mu$ is the easiest to tune in experimental setup (through the variation of the gate voltage). $\alpha$ and $\Delta$ are the material parameters, which are hard to tune experimentally. $\theta$ is the dynamic parameter (in the sense it depends on other parameters $\alpha$, $\Delta$, $\mu$); it should take the value to decrease the free energy of the system. With these considerations, it is interesting to investigate: (i) how $\theta$ (for fixed $\alpha$ and $\Delta$) dynamically reacts to a change in $\mu$; (ii) can the tuning of $\mu$ be used to drive the system into and out of the topologically non-trivial phase?

To answer these questions, we numerically calculate the dependence of the free energy on all the parameters of the system. This is done by diagonalizing and integrating the Hamiltonian, Eq. (\ref{eq:ham-mom}), over the whole BZ; it should be kept in mind that for a non-zero $\alpha$ the BZ gets folded $N=2\pi/\theta$ times (the periodicity of the lattice). In Fig. \ref{fig:min-ener-plt} we plot the dependence of free energy on $\theta$ for series of values of $\mu/t$ at $\alpha/t=\Delta/t=0.2$. In this plot the $\theta$ corresponding to the minimum energy is marked by down triangle. It shows that for $\mu/t<0.8$ the minimum free energy occurs for $\theta=\pi$ (the AFM order). However with increase in $\mu/t$ the minimum energy $\theta$ shifts away from $\theta=\pi$. It means there may be some window of values of the parameters $\mu$, $\alpha$ and $\Delta$, where topological phase might appear thermodynamically.

\begin{figure}[tbh]
  \centering
  \includegraphics[width=0.48\textwidth]{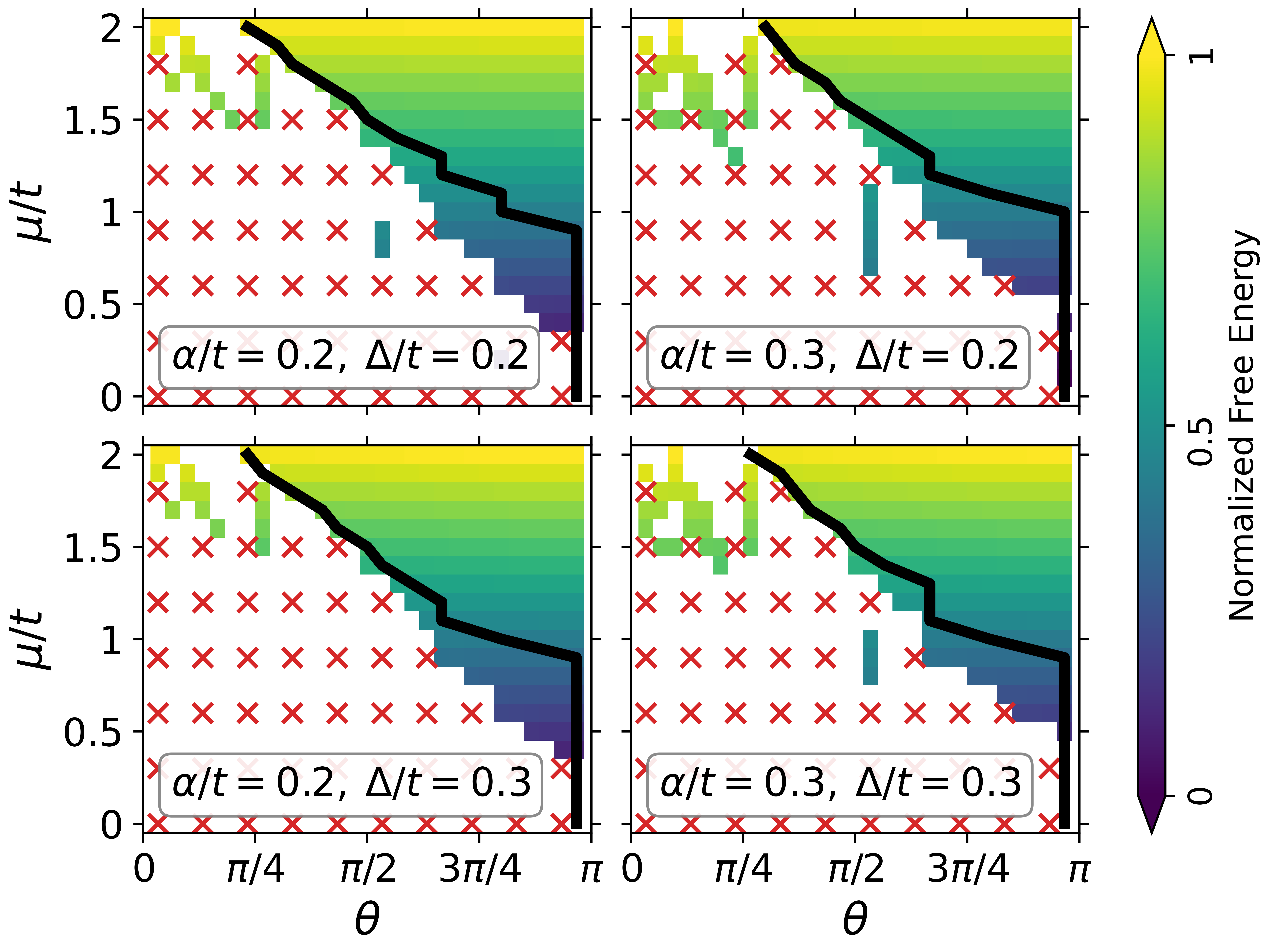}
  \caption{In each subfigure the dependence of the free energy on dynamical parameter $\theta$ (\emph{x}-axis) and external parameter $\mu$ (\emph{y}-axis) is shown for $\alpha/t=0.2,0.3$ and $\Delta/t=0.2,0.3$. The topologically trivial regions is marked by crosses ($\boldsymbol{\times}$). The free energy is shown only for topologically non-trivial regions. The bold black line represent the evolution of the $\theta$ corresponding to the minimum free energy as $0<\mu/t<2$ is varied. If the minimum free energy $\theta$ (bold black line) has values away from $\theta \approx 0, \pi$ and outside the topologically trivial region then the system is in topologically non-trivial state.}
  \label{fig:seq-dynamic-theta}
\end{figure}
To understand better the dependence of thermodynamically occurring topologically non-trivial phase on the values of the parameter, in Fig. \ref{fig:seq-dynamic-theta} we plot the dependence of the free energy on the dynamic parameter $\theta$ and on the external parameter $\mu/t$ for $\alpha/t=0.2,0.3$ and $\Delta/t=0.2,0.3$. The free energy is shown only for those $\theta,\mu$ where the topologically non-trivial phase occurs (shown by color gradients); the region where trivial topological phase occurs the free energy is not shown, and marked by crosses ($\boldsymbol{\times}$). The phase is topologically non-trivial if the following two conditions are satisfied: (i) $\theta$ should be contained within the range $\theta \in \left( 0, \pi \right)$, (ii) the $\mathbb{Z}_{2}$ topological invariant should be $Q=-1$; calculated using Eq. (\ref{eq:z-2-invariant}). In Fig. \ref{fig:seq-dynamic-theta} we also show the values of $\theta$ corresponding to the minimum free energy ($\theta_{\text{min}}$, bold black line); minimum $\theta$ is chosen by considering both trivial and non-trivial phase energy. The answer to the question, whether or not it is possible to use $\mu$ for driving the system into or out of the topologically non-trivial phase, is affirmative, provided the following conditions are satisfied. First, $\theta_{\text{min}}$ lies in the region where the aforementioned topological conditions are satisfied. Second, a non-stringent condition, $\theta_{\text{min}}$ does not lie in the vicinity of the boundary of the topological and trivial phase. This latter condition is added to make sure that a small perturbation of the system should not affect the topological properties \footnote{Although, by definition the topological properties of the system should not be prone to perturbation, however, here we are talking about the perturbation around the limiting values of the parameters.}. As an example in Fig. \ref{fig:seq-dynamic-theta} for $\alpha/t=0.2$, $\Delta/t=0.2$ when $0< \mu/t \lessapprox 0.8$ the $\theta_{\text{min}} \approx \pi$, hence, the system is in trivial phase. For $0.8 \lessapprox \mu/t \lessapprox 1.5$ the $\theta_{\text{min}} \not \approx 0, \pi$, and inside the region where topological conditions are satisfied; therefore the system is in a topological phase. However, for $\mu/t \gtrapprox 1.5$ the $\theta_{\text{min.}}$ is at the boundary of both a trivial and a non-trivial topological phase, and we can no longer guarantee the stability of such topological phase.

One of the peculiar things in these figures is the presence of topologically non-trivial islands inside the trivial phase, e.g. when $\alpha/t=0.3$ and $\Delta/t=0.2$ such topological phase is present for $\pi/8 \lessapprox \theta \lessapprox \pi/2$ and $1.5 \lessapprox \mu/t \lessapprox 2$. This happens due to the re-entrant nature of the topological phase.

\section{Proposal for experimental realization}
\label{sec:heter-exper-real}
To observe the predicted effect three ingredients are necessary: (i) a 1D strongly correlated electronic channel with an extended \emph{s}-wave superconducting order parameter and non-zero RSOC, (ii) a mechanism to tune chemical potential, (iii) a mechanism to detect MFs. One of the possible way for satisfying the first requirement is to use the heterostructures involving Moire lattices and vdW materials; the schematic is shown in Fig. \ref{fig:scheme-device-moire}.

The canonical way to induce strong correlation is to use materials with localized $d$ and $f$ electrons \footnote{Due to localized nature of the $d$ and $f$ electrons the corresponding band width $W$ is narrow, quenching the kinetic energy. The localized electrons now feel strong coloumb forces of nearby electrons giving rise to strong electronic correlation phenomenons.}. Another more recent strategy is to use so called ``flatband engineering'' \cite{balents-2020-super-stron}; in this case kinetic energy is quenched by destructive interference in frustrated lattice \cite{derzhko-2015-stron-correl,ye-2024-hoppin-frust} or increase in unit cell by moire potentials \cite{leykam-2018-artif-flat}. The vdW transition metal dichalcognides \cite{novoselov-2016-mater-van} and vdW magnets \cite{yang-2021-van-der} provide suitable platforms for inducing strong correlation through all of the three aforementioned strategies \cite{balents-2020-super-stron,liu-2023-fabric-energ,checkelsky-2024-flat-bands}.

Parallel (Quasi-)1D Moire patterns (electronic channels) on a 2D layered vdW materials can be produced by stretching or straining \cite{drozdz-2024-quasi-moire,sinner-2023-strain-induc,santos-2021-super-twist}. Single electronic channels can be isolated by cutting parallel 1D Moire patterns by lasers or atomic force microscope (see \cite{lau-2022-reprod-fabric} and references therein). We assume that the electronic channels are long enough (along $x$-axis) so that electronic state quantization along the length can be ignored, and thin enough (along $y$-axis) so that the 1D sub-bands are well separated on the relevant energy scale. To induce extended \emph{s}-wave superconductivity in these isolated electronic channels one can use iron-based SC, e.g. Fe(Se,Te), Pr doped CaFe$_{2}$As$_{2}$ ($T_{c}\approx 49$K \cite{lv-2011-unusual-super}), or oxyarsenide Sm[O$_{1-x}$F$_{x}$]FeAs ($T_{c} \approx 55$K \cite{zhi-an-2008-super-at}). Proximity induced superconductivity has been confirmed in Fe(Se,Te) (see references within \cite{[][]zhu-2023-proxim-effec}). Second requirement is satisfied through gate voltage; by varying the gate voltage one can tune the chemical potential. In Fig. \ref{fig:scheme-device-moire} super gates (orange gates) satisfy this function.

The third requirement can be satisfied by biasing the 1D channel and measuring the differential conductivity (\emph{dI/dV}). Back gates (blue gates in Fig. \ref{fig:schematics}) are used for biasing the electronic channel. Tunnel gates (pink gates) are used to induce tunnel potential; it is needed to control the electron flow. The tunneling spectroscopy is found by measuring \emph{dI/dV} increasing the back gate bias voltage ($V$) from $-V_{0}$ to $+V_{0}$ at constant tunnel gate voltage and super gate voltage. We expect a peak in \emph{dI/dV} in the presence of MF.

Apart from proposed vdW setup in Fig. \ref{fig:schematics}, one can use the already widely available InAs setup \cite{flensberg_2021_EngineeredPlatforms_NatRevMater,frolov-2020-topol-super}, because recently it was predicted that in depleted InAs NW the \emph{e-e} correlation is strong enough \cite{heedt-2017-signat-inter,sato-2019-stron-elect} for emergence of the strongly correlated behaviours. The only difference between the InAs setup in Ref. \cite{flensberg_2021_EngineeredPlatforms_NatRevMater} and the setup to realize the proposed physics is the presence of an extra gate to deplete InAs NW and absence of the magnetic field. Experimentally, analogous heterostructures with gates for depleting InAs NW have already been synthesized in Ref. \cite{heedt-2017-signat-inter}; in fact the re-entrant conductance has been observed in their heterostructures for non-zero $\alpha$ and zero magnetic field. There is no superconductivity in the heterostructure of Ref. \cite{heedt-2017-signat-inter}, however, using this as the starting point one can synthesize the heterostructure with iron-based SC.

The presence of the zero bias \emph{dI/dV} peak does not always mean emergence of the MF. The zero bias peak can also appear due to Andreev bound states \cite{prada_2020_AndreevMajorana_NatRevPhys}. Hence, several other non-local MF detection methods are also proposed \cite{zhang_2019_NextSteps_NatCommun}. The proposed heterostructures (both vdW and InAs based) can be modified accordingly for non-local detection of the MF.

\section{Possible strategies for generation and control of spiral spin structure}
\label{sec:gener-spir-spin}
\begin{figure}[tbh]
  \centering
  \includegraphics[width=0.45\textwidth]{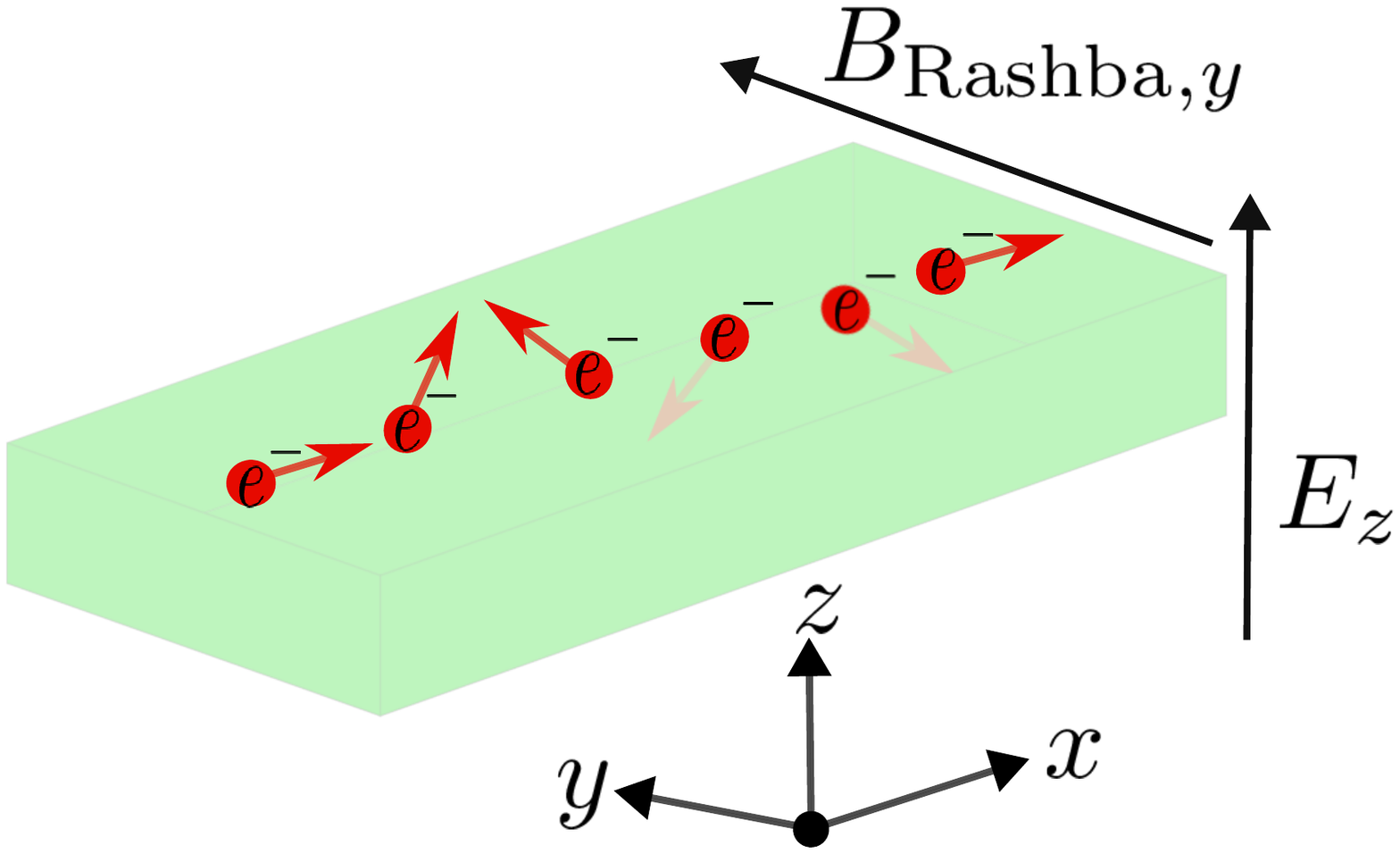}
  \caption{Generation and control of the spiral spin structure due to Rashba Field. The conducting electrons enters into the materials along $x$-axis (from left to right), an external electric field is applied along $z$-axis, the resulting Rashba field $\mathbf{B}_{\text{Rashba}}$ is along $y$-axis. If the spin of the electrons is not aligned along the $\mathbf{B}_{\text{Rashba}}$ then it precesses around $\mathbf{B}_{\text{Rashba}}$. In figure the initial spin (arrows) of the electron is along $x$-axis, hence it precesses around $y$-axis. Electric field $E_{z}$ is used to control the direction and intensity of the $\mathbf{B}_{\text{Rashba}}$ \cite{manchon_2015_NewPerspectives_NatureMater}.}
  \label{fig:spiral-Rashba}
\end{figure}

The discussion in previous sections heavily relied on the spiral spin structure in the system, therefore, it is desirable to have methods to generate and control them. One of the feasible and easiest way is to use Rashba field $\left(\mathbf{B}_{\text{Rasbha}}  \right)$ for this purpose. Electron travelling in an electric field feel a magnetic field proportional to the electric field. The effective magnetic field couples with the spin of the electron which gives rise to the spin-orbital effects. In systems where structural inversion symmetry is broken the spin-orbital Hamiltonian can take a simplistic form $H_{SO} \sim \alpha \left[ \mathbf{z} \times \mathbf{k} \right] \cdot \sigma$ (assuming electric field $E_{z}$ is along $z$-axis). Here $\alpha$ is the Rashba parameter which can be controlled by the applied gate electric field \cite{nitta-1997-gate-contr,schultz-1996-rashb-spin,koo-2009-control,liang-2012-stron-tunin,takase-2017-highl-gate}; $\mathbf{z}$ is the unit vector along $z$-axis; $\mathbf{k}$ is the electron momentum; $\sigma$ is the vector of Pauli spin operator. Consequently if electron flows along $x$-axis then it feels a magnetic field (Rashba field $\mathbf{B}_{\text{Rasbha}}$) along $y$-axis. For a quasi-one dimensional system the Rashba field approximates to $\mathbf{B}_{\text{Rashba}} \approx \alpha k_{F}/g\mu_{B}$. Here $k_{F}$ is the Fermi momentum; $g$ is the electron $g$-factor; $\mu_{B}$ is the Bohr magneton. If the spin of the conducting electron is not aligned along the Rashba field then it starts to precesses around the corresponding field. Hence if spin of the conducting electron is aligned along $x$-axis then it precesses around $y$-axis, and an out of plane spiral spin structure is generated. It is explained in Fig. \ref{fig:spiral-Rashba}.

$\mathbf{B}_{\text{Rashba}}$ also plays the role in controlling the spiral modulation angle, as the rate of precession (frequency) depends on its intensity. One can approximate spatial angle modulation $\Delta \theta \approx |\mathbf{B}_{\text{Rashba}}|/v_{F}a$; here $v_{F}=\partial \varepsilon/\partial k|_{k = k_{F}}$ is the Fermi velocity which is band structure dependent; a is the lattice constant. We can see that higher the $|\mathbf{B}_{\text{Rashba}}|$, higher will be spatial modulation. As $\mathbf{B}_{\text{Rashba}}$ is a function of $\alpha$, which in turn can be changed by the external gate voltage, therefore, the electric field can control both spiral modulation angle and direction of the spiral modulation (provided structural inversion symmetry is broken along the direction of the electric field).

InAs based heterostructure (discussed in previous section) provide useful setup for this purpose. There are mainly two reasons for this. Firstly, Rashba physics is effective for narrow gap semiconductor like InAs. Therefore most of the experiments where electrical control of the $\alpha$ is shown is done on InAs based heterostructures \cite{nitta-1997-gate-contr,schultz-1996-rashb-spin,koo-2009-control,liang-2012-stron-tunin,takase-2017-highl-gate,scheruebl-2016-elect-tunin}. Secondly, in the same InAs NW strong \emph{e-e} can be induced through depletion \cite{sato-2019-stron-elect,heedt-2017-signat-inter}, hence satisfying one of the major requirement of the Hamiltonian. Essentially, we propose to use the InAs platforms described in \cite{flensberg-2021-engin-platf} as the base (of course with extended $s$-wave superconductor); further add extra gates for depletion of InAs \cite{heedt-2017-signat-inter} and electrical control of $\mathbf{B}_{\text{Rashba}}$ \cite{scheruebl-2016-elect-tunin}.

Most of the widely studied promising vdW materials, e.g. transition metal dichalcogenides, trihalides or phosphorus trichalcogenides, are either centrosymmetric (1T) or non-centrosymmetric (2H) in pristine condition (in 2H structure the planar mirror symmetry is present). In these materials hexagonal lattice of metallic elements are sandwiched between two layers of chalcogens (S, Se, Te) or halides (Cl, Br, I), which don't break the structural inversion symmetry (the must have for Rashba physics to appear). Hence, one need to break the 1T and 2H symmetries. Usually it is done by implanting the chalcogen atoms of the upper (or lower if possible) atoms by other suitable atoms. These type of materials are known as Janus 2D materials \cite{ju-2020-two-dimen}. The RSOC has been observed in these materials \cite{hu-2018-intrin-anisot} and can be controlled by the electric field and strain \cite{patel-2022-elect-field}. Besides strong correlation is predicted to be present in these materials \cite{angeli-2022-twist-janus}. Therefore one can use the vdW heterostructure as proposed in Fig. \ref{fig:scheme-folding-of-BZ} with Janus materials, and use electric field to control the Rashba parameter.

Another method to generate spiral spin structure is through Dzyaloshinskii-Moriya interactions (DMI) \cite{fert-2023-early,kuepferling-2023-measur-inter}. DMI is an anti-symmetric interaction which forces the neighbouring spins to align perpendicular to each other ($ \propto D_{ij} S_{i} \times S_{j}$; $D_{ij}$ is the DMI parameter; $S_{i}$ and $S_{j}$ are the neighbouring spin). In conjunction with Heisenberg interaction --- which forces the neighbouring spins to align either parallel (FM) or anti-parallel (AFM) to each other ($\propto \pm J S_{i} \cdot S_{j}$; $\pm J$ is the exchange interaction) --- the DMI gives rise to spiral spin structure. In fact in number of vdW heterostructures the spiral structure has already been observed \cite{meijer_2020_ChiralSpin_NanoLett,walsem-2020-layer-effec,du-2023-strain}. $D_{ij}$ can be manipulated by electric field \cite{kammerbauer-2023-dzyal-moriy} or by strain \cite{gusev-2020-manip-dzyal}. As intensity of $D_{ij}$ is directly related to the inclination of spin away from each other, by manipulating $D_{ij}$ one can control the spiral spin structure.

On a side note, we believe, although vdW based heterostructures are promising for generating spiral spin field, however, the InAs based heterostructure are more feasible as the techniques for their synthesis and measurement are readily available, especially the strong correlation can be readily induced them through depletion \cite{sato-2019-stron-elect,heedt-2017-signat-inter}.

\section{In-plane magnetic field and its effect on Hamiltonian}
\label{sec:contr-peri-hamilt}
It is interesting to investigate how the Hamiltonian will change if the magnetic field is not along $z$-axis but along arbitrary direction. Physically, one possible application of this analysis will be the case where magnetic field is applied along the plane of spiral spin structure. The magnetic field along arbitrary direction reads
\begin{equation}
  \label{eq:mag-field}
   \mathbf{B} = B_{x} \hat{x} + B_{y} \hat{y} + B_{z} \hat{z}.
 \end{equation}
 Here $B_{x}$, $B_{y}$ and $B_{z}$ are the $x$, $y$ and $z$ component of the magnetic field respectively; $\hat{x}$, $\hat{y}$, and $\hat{z}$ are the unit vectors along $x$, $y$ and $z$ directions. The Hamiltonian of a 1D SC NW, Eq. (\ref{eq:basic-ham}), under an externally applied magnetic field along arbitrary direction reads
 \begin{equation}
  \label{eq:basic-ham-with-arb-mag}
  \begin{aligned}
    H_{B}=&-c^{\dagger}_{i\sigma} \left( t \taus[0] + \alpha \taus[y] \right) c_{i+1\sigma'}
        -\imath c^{\dagger}_{i\sigma}\Delta \taus[y]c^{\dagger}_{i+1 \sigma'}+\text{H.c.}\\
      &\quad + c^{\dagger}_{i\sigma} \left( \mu \taus[0] +  B_{x} \taus[x] +  B_{y} \taus[y] + B_{z} \taus[z] \right) c_{i+1\sigma'}\\
    & \qquad + U\sum_in_{i\uparrow}n_{i\downarrow}.
  \end{aligned}
\end{equation}
In comparison to Eq. (\ref{eq:basic-ham}) additional terms containing $B_{x}$ and $B_{y}$ appears in Eq. \eqref{eq:basic-ham-with-arb-mag}. For simplicity one can rewrite Eq. (\ref{eq:basic-ham-with-arb-mag}) as:
 \begin{equation}
  \label{eq:basic-ham-with-arb-mag-trans}
  \begin{aligned}
    H_{B}=&-c^{\dagger}_{i\sigma} \left[ t \taus[0] + \left( \alpha - B_{y}/2  \right) \taus[y] - \left(B_{x}/2\right) \taus[x] \right] c_{i+1\sigma'}\\
      &\quad -\imath c^{\dagger}_{i\sigma}\Delta \taus[y]c^{\dagger}_{i+1 \sigma'}+\text{H.c.}\\
      &\qquad + c^{\dagger}_{i\sigma} \left( \mu \taus[0] + B_{z} \taus[z] \right) c_{i+1\sigma'} + U\sum_in_{i\uparrow}n_{i\downarrow}.
  \end{aligned}
\end{equation}
Now the $B_{y}$ and $B_{x}$ are absorbed into the first term. In terms of Hubbard operator Eq. (\ref{eq:basic-ham-with-arb-mag-trans}) will be same to the Eq. (\ref{eq:strong-ham}) apart from the term containing hopping potential, which becomes
\small
$$\left(t \tau_{0}^{\sigma\sigma'} + \alpha \tau_{y}^{\sigma\sigma'} \right) \to  \left[ t \taus[0] + \left( \alpha - B_{y}/2  \right) \taus[y] - \left(B_{x}/2\right) \taus[x] \right].$$
\normalsize
After the $su(2|1)$ coherent state transformation (see Sec. \ref{sec:model} and \ref{sec:su21} for derivation) Eq. (\ref{eq:basic-ham-with-arb-mag-trans}) reads
 \begin{equation}
   \label{eq:effectiv-ham-trans-arb-B}
   \begin{aligned}
     &H_{\text{eff},B}(z, \xi) =\\
     &\quad - \sum\limits_{i} \xi_{i+1} \bar{\xi}_{i} \left[t \: a_{i} + \imath \left( \alpha - B_{y}/2  \right) \alpha_{i} - \left( B_{x}/2 \right) \gamma_{i} \right] \\
     &\qquad - \Delta\sum\limits_{i}\bar{\xi}_{i+1}\bar{\xi}_{i}\: \Delta_{i} + \text{H.c.} + \sum\limits_{i} \xi_{i} \bar{\xi}_{i} \left(\mu + B_{z}\beta_{i}\right).
   \end{aligned}
\end{equation}
The definitions of the parameters $a_{i}$, $\alpha_{i}$, $\Delta_{i}$, and $\beta_{i}$ are same as in Eq. (\ref{eq:def-aij}); the new term $\gamma_{i}$ is defined as:
\begin{equation}
  \label{eq:gamma-i}
  \gamma_{i} \equiv \frac{\bar{z}_{i+1} + z_{i}}{\sqrt{\left( 1 + \left| z_{i+1} \right|^{2}\right) \left( 1 + \left| z_{i} \right|^{2}\right)}}.
\end{equation}

For a spinon field having only $xy$ component, Eq. (\ref{eq:spin-config-spiral-xy}), the $a_{i}$, $\alpha_{i}$, $\Delta_{i}$, and $\beta_{i}$ will be same as Eq. (\ref{eq:spiral-spin-aij}); the new term
\begin{equation}
  \label{eq:gamma-i-arb-B}
  \begin{aligned}
   &\gamma_{i} = \mathrm{e}^{-\imath \left(\theta_{i+1}-\theta_{i}\right)/2}  \cos \frac{\left(\theta_{i+1}+\theta_{i} \right)}{2}.
  \end{aligned}
\end{equation}
In case of spiral configuration ($\theta_{i+1}-\theta_{i} = \theta$) the effective Hamiltonian, Eq. (\ref{eq:effectiv-ham-trans-arb-B}), will transform to
\begin{equation}
  \label{eq:effectiv-ham-spiral-spin-arb-B}
  \begin{aligned}
    H_{\text{eff},B}=-\sum\limits_{i} \xi_{i+1} \bar{\xi}_{i} \: \tilde{t}_{i,B} - \Delta \sin \frac{\theta}{2}\sum\limits_{i}\bar{\xi}_{i+1}\bar{\xi}_{i} + \text{H.c.} + \mu\sum\limits_{i} \xi_{i}\bar{\xi}_{i},
  \end{aligned}
\end{equation}
where,
\begin{equation*}
  \tilde{t}_{i,B} = t \cos \frac{\theta}{2} + \left( \alpha-B_{y}/2  \right) \sin \left( \theta_{i} + \frac{\theta}{2} \right) - \left(B_{x}/2  \right) \: \cos \left( \theta_{i} + \frac{\theta}{2} \right).
\end{equation*}
Comparing Eq. \eqref{eq:effectiv-ham-spiral-spin-arb-B} and (\ref{eq:effectiv-ham-spiral-spin}) we observe that both the Hamiltonians are essentially same and have the periodic nature, with a small difference, the periodicity in Eq. (\ref{eq:effectiv-ham-spiral-spin-arb-B}) is a function of the in-plane magnetic field through $\tilde{t}_{i,B}$.

\begin{figure}[tbh]
  \centering
  \includegraphics[width=0.48\textwidth]{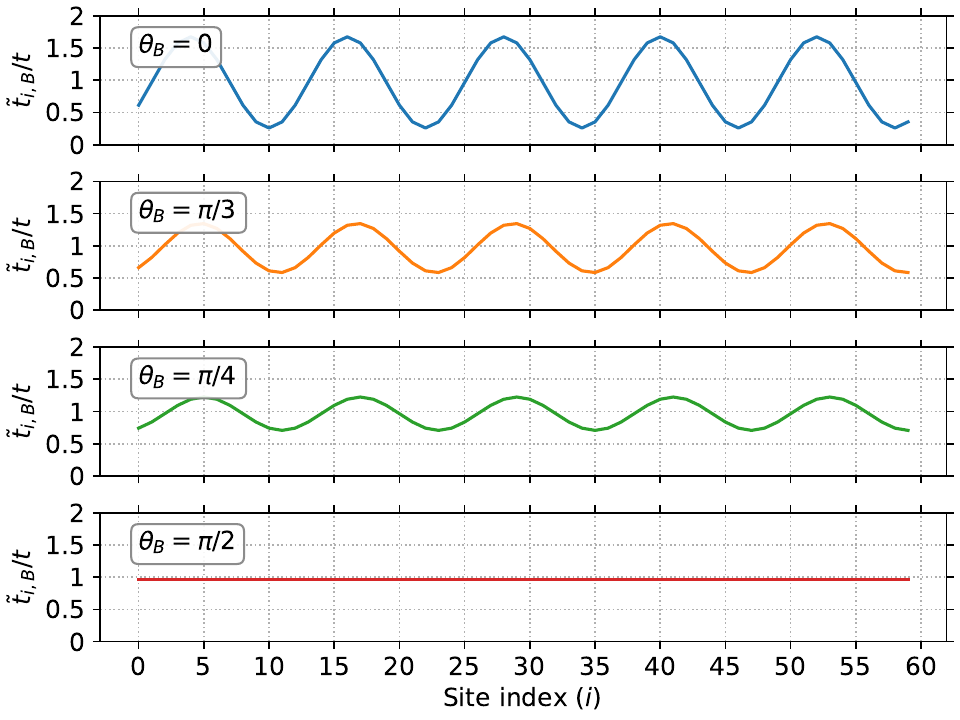}
  \caption{Dependence of the effective hopping term $\tilde{t}_{i,B}$ in Eq. (\ref{eq:effectiv-ham-spiral-spin-arb-B}) on the in-plane magnetic field components $B_{x}=\left(|\mathbf{B}|/t \right) \cos \theta_{B}$ and $B_{y}=\left( |\mathbf{B}|/t \right) \sin \theta_{B}$. The RSOC is $\alpha/t=0.5$; intensity of magnetic field is $|\mathbf{B}|/t=1$; and $\theta_{B}$ is the anticlockwise azimuthal angle from positive $x$-direction. The $x$-axis of the figure represent the site index, and $y$-axis represent the $\tilde{t}_{i,B}/t$. The hopping factor $\tilde{t}_{i,B}$ becomes constant when $\theta=\pi/2$ and $|\mathbf{B}|=2\alpha$ as in the bottom figure.}
  \label{fig:in-plane-hop}
\end{figure}
The in-plane magnetic field presents a way to control the sub-band energy gap which emerges due to periodicity of $\tilde{t}_{i,B}$. In Eq. (\ref{eq:effectiv-ham-spiral-spin}) the sub-band gap was controlled by the $\alpha$; with increase in $\alpha$ the sub-band gap increases (see Fig. \ref{fig:ener-spect-alpha-dep}). In Eq. (\ref{eq:effectiv-ham-spiral-spin-arb-B}) the same role is being played by the the terms $\left(\alpha-B_{y}/2 \right)$ and $B_{x}/2$. It should be noted that the periodicity of the whole term $\left( \alpha-B_{y}/2  \right) \sin \left( \theta_{i} + \frac{\theta}{2} \right) - \left(B_{x}/2  \right) \: \cos \left( \theta_{i} + \frac{\theta}{2} \right)$ is independent of the parameters $\alpha$, $B_{x}$ and $B_{y}$; these parameters only affect the amplitude of the $\tilde{t}_{i,B}$ as shown in Fig. \ref{fig:in-plane-hop}. Another interesting fact which can be observed in Fig. \ref{fig:in-plane-hop} is the absence of periodicity in $\tilde{t}_{i,B}$ for $\theta_{B}=\pi/2$. It happens only when two special conditions are satisfied: (i) the intensity of the magnetic field is $|\mathbf{B}|=2 \alpha$, (ii) direction of the applied magnetic field is along $y$-axis. Therefore by rotating the magnetic field (of intensity $|\mathbf{B}|=2 \alpha$) on the $xy$ plane one can open and close the induced sub-band energy gaps; it also results in emergence and disappearance of the re-entrant nature.

Another physical consequence due to the form of the $\tilde{t}_{i,B}$ is, when $\theta_{B}=\arcsin \left( 2\alpha/|\mathbf{B}| \right)$ the periodicity of $t_{i,B}$ becomes clean sinusoidal. Essentially when the condition for $\theta_{B}$ is satisfied the term containing $(\alpha-B_{y}/2)$ vanishes; there remains only the term $\left( B_{x}/2  \right) \cos \left( \theta_{i} + \theta/2 \right)$. If the condition for $\theta_{B}$ is not satisfied, then both the
$\left( \alpha -B_{y}/2 \right) \sin \left( \theta_{i} +\theta/2 \right)$ and $\left(B_{x}/2  \right) \cos \left( \theta_{i} +\theta/2 \right)$ terms survives; summation of which is a mixed sinusoidal. Both these cases affects the sub-band energy gap.

In-plane magnetic field provides interesting external control to change the properties of the system. Qualitatively, the system will show the same re-entrant topological behaviour as discussed in Sec. \ref{sec:phase-diagram}. The only difference is now through the rotation of the magnetic field we can control sub-band energy gaps, and even completely close it (when $|\mathbf{B}|=2\alpha$ and $\theta_{B}=\pi/2$). Hence all the proposals and schemes for heterostructures discussed in Sec. \ref{sec:heter-exper-real} and \ref{sec:gener-spir-spin} are also applicable here. The only difference is one need to add an extra in-plane magnetic field.

\begin{figure}[tbh]
  \centering
  \includegraphics[width=0.45\textwidth]{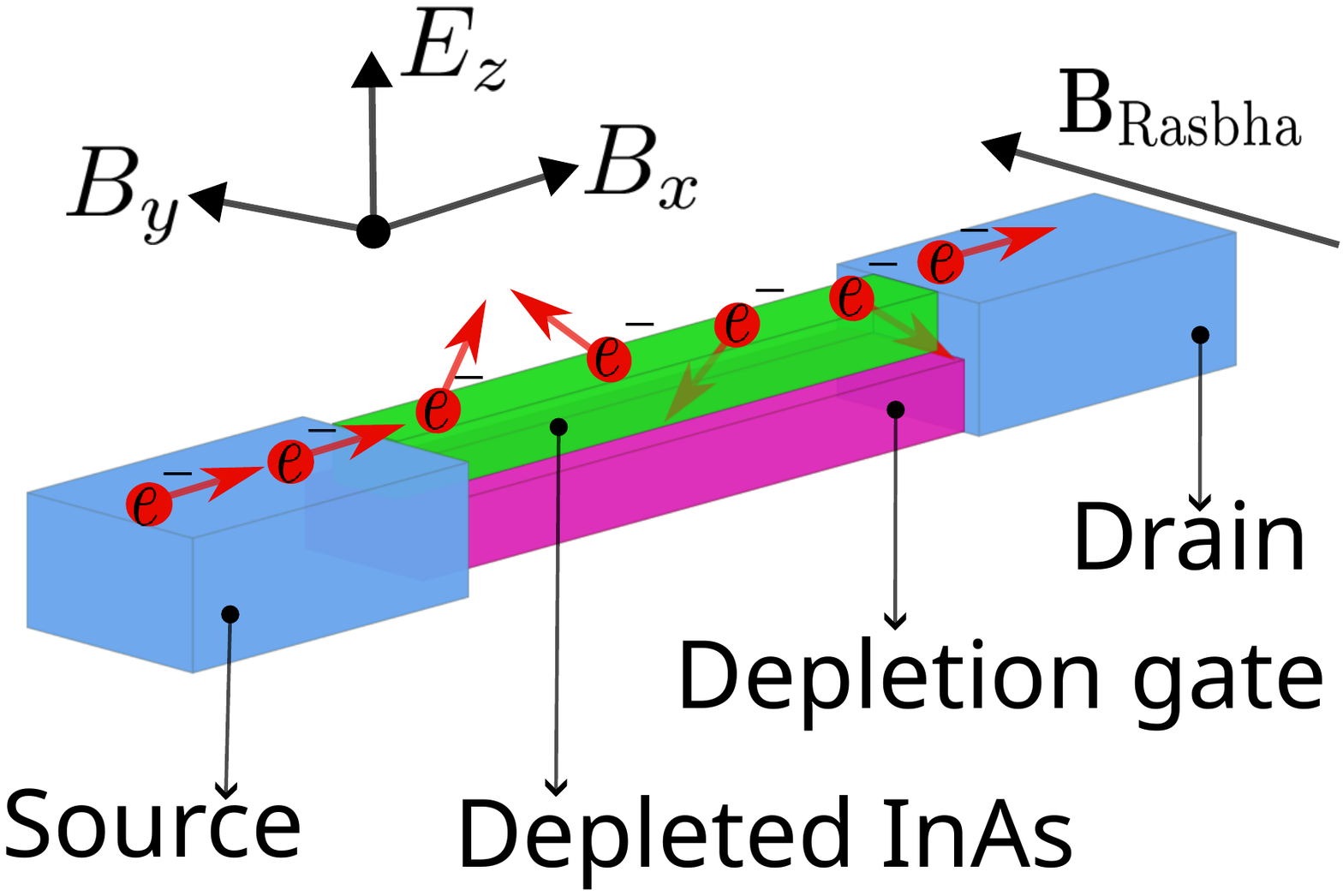}
  \caption{Color online: Schematics of the heterostructure for conductance experiment using InAs under in-plane magnetic field ($B_{x},B_{y}$) and out of plane electric field ($E_{z}$). The Rashba field $\mathbf{B}_{\text{Rashba}}$ is along $y$-axis. Th InAs NW (green) is depleted using depletion gate (pink). Electrons (red balls) with spin (arrows) along $x$-direction travels from source to drain. The spin of the electron precesses around the $y$-axis generating a spiral spin structure. By varying externally applied field ($B_{x}$, $B_{y}$ and $E_{z}$) one can expect interesting re-entrant conductance behaviour. The conductance experiment on depleted InAs without the in-plane magnetic field is done in Ref. \cite{heedt-2017-signat-inter}.}
  \label{fig:scheme-InAs}
\end{figure}

Although heterostructures with superconductors are good systems to investigate the effect of the in-plane magnetic field, however, the systems without superconductivity provide better opportunity (as one does not need to incorporate superconductivity in them). The Hamiltonian for non-superconducting systems can be found by putting $\Delta=0$ in Eq. \eqref{eq:effectiv-ham-trans-arb-B}. We propose a heterostructure as shown in Fig. \ref{fig:scheme-InAs} using depleted InAs semiconducting NW, where strong \emph{e-e} correlation appears \cite{sato-2019-stron-elect,heedt-2017-signat-inter}. Basically the idea is to take a depleted InAs semiconducting NW as in Ref. \cite{heedt-2017-signat-inter}, and apply an in-plane magnetic field (along $xy$-plane) and an out-of plane electric field (along $z$-axis). The in-plane magnetic field is needed to control the amplitude of the sinusoidally varying $t_{i,B}$ (difference between maximum and minimum). The out of plane electric field (perpendicular to the NW axis) is needed to control the direction and intensity of the Rashba field $\mathbf{B}_{\text{Rashba}}$ (explained in Sec. \ref{sec:gener-spir-spin}). $\mathbf{B}_{\text{Rashba}}$ is always directed perpendicular to the conduction electron velocity and external electric field, and the spin of the electrons rotates around $\mathbf{B}_{\text{Rashba}}$ (see Fig. \ref{fig:spiral-Rashba}). Therefore direction of the spin spiral (axis around which spin rotates) can be varied by changing the direction of the electric field. The intensity of $|\mathbf{B}_{\text{Rashba}}|$ controls the rate of precession of electronic spin, which results in controlling the modulation angle $\theta$. One can approximate the modulation angle $\theta \sim \alpha k_{F}/v_{F}$ (see Sec. \ref{sec:gener-spir-spin}); $k_{F}$ and $v_{F}$ are the Fermi momentum and velocity respectively. Moreover $|\mathbf{B}_{\text{Rashba}}|$ depends on Rashba parameter $\alpha$, which can be varied by external electric field \cite{nitta-1997-gate-contr,schultz-1996-rashb-spin,koo-2009-control,liang-2012-stron-tunin,takase-2017-highl-gate,scheruebl-2016-elect-tunin}. Interesting scenario appears as now both $\alpha$ and $\theta$ in $\tilde{t}_{i,B}$ is controlled by the external electric field. By varying $\mathbf{B}$ and $E_{z}$ one can find interesting behaviour. Similar experiment without in-plane magnetic field is done in Ref. \cite{heedt-2017-signat-inter}. In depth theoretical investigation of this system is out of the scope of this article, hence will be discussed in another article.

\section{Conical spin structure and emergence of the pair density wave}
\label{sec:conic-spin-struct}
\begin{figure}[tbh]
  \centering
  \includegraphics[width=0.45\textwidth]{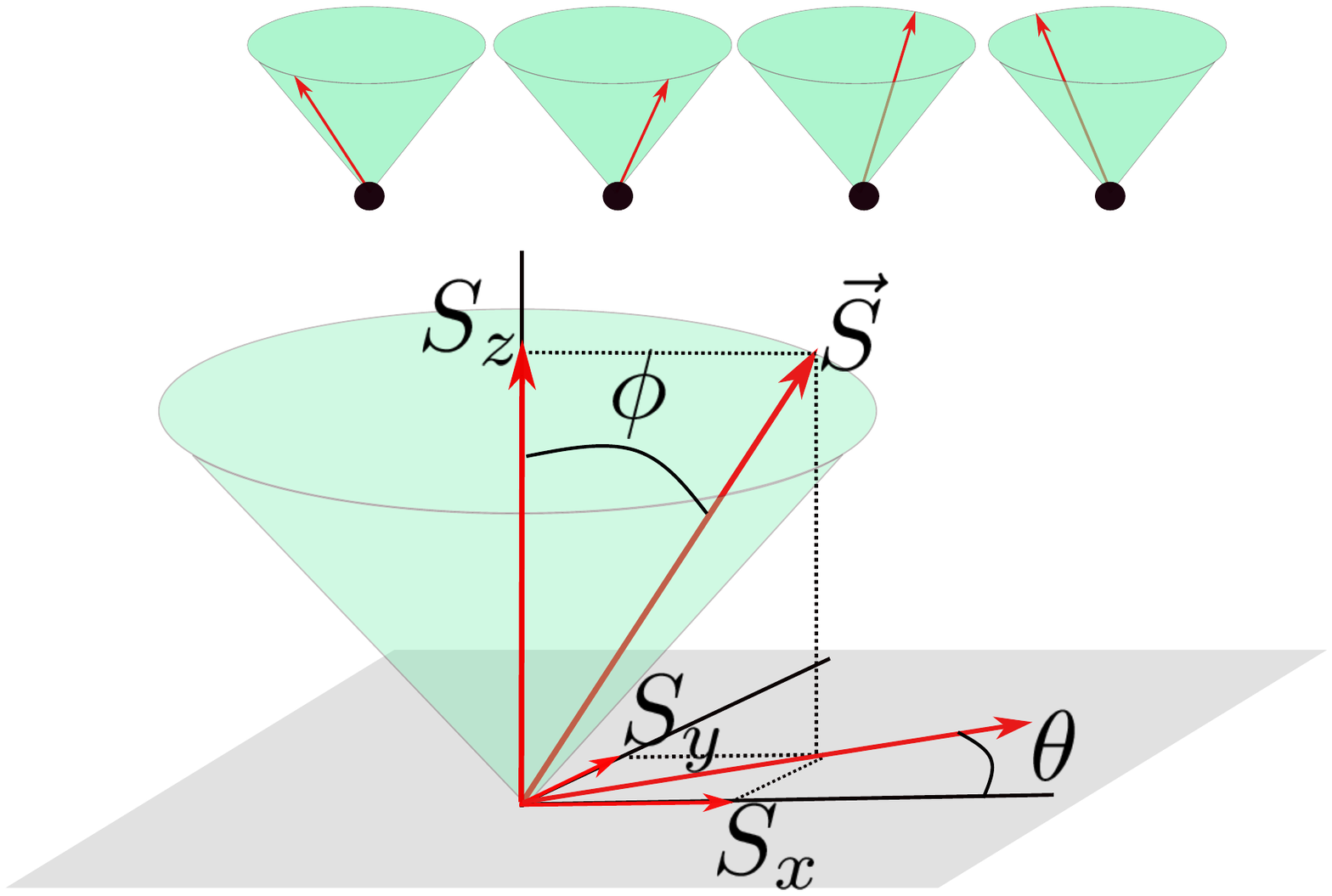}
  \caption{Representation of spiral spin structure. $S^{x}$, $S^{y}$ and $S^{z}$ are the $x$, $y$, and $z$ component resepctively. $\theta$ is the polar angle and $\phi$ is the azimuthal angle. With the $z$-component of the spin $S^{z}$ fixed it can be thought of as rotating on the surface of the cone.}
  \label{fig:spiral-spin}
\end{figure}
Conical spin structure is widely observed in magnetic materials, hence it is interesting to see its effect on the overall Hamiltonian. The spin structure is shown in Fig. \ref{fig:spiral-spin}, and can be represented as:
\begin{equation}
  \label{eq:spin-config-conical}
  \vec{S}_{i} = \left( S_{i}^{x}, S_{i}^{y}, S_{i}^{z} \right) = \frac{1}{2}\left( \sin \phi \cos \theta_{i}, \sin \phi \sin \theta_{i}, \cos \phi  \right).
\end{equation}
Here $\phi$ is the polar angle and $\theta$ is the azimuthal angle. One should note that $\phi$ is spatially independent (constant). The corresponding spinon $c$-number can be found by using Eq. (\ref{eq:z-real-imag-polar}):
\begin{equation}
  \label{eq:conical-zi}
  z_{i}=\left(\frac{\sin \phi}{1+\cos \phi}   \right)\mathrm{e}^{\imath \theta} \equiv \Gamma \mathrm{e}^{\imath \theta}.
\end{equation}
Using Eq. \eqref{eq:conical-zi} the spinon fields $a_{i}$, $\alpha_{i}$, $\Delta_{i}$ and $\beta_{i}$ can be found from Eqs. (\ref{eq:spiral-spin-aij}):
\begin{widetext}
\begin{equation}
  \label{eq:conical-spin-aij}
  \begin{aligned}
    &a_{i} = \exp {-\imath \left[\left(\frac{\theta_{i+1}-\theta_{i}}{2}\right) - \arctan \left[ \left( \frac{1-\Gamma^{2}}{1+\Gamma^{2}} \right) \: \tan \frac{\left(\theta_{i+1}-\theta_{i}\right)}{2} \right]\right]}
      \times \sqrt{ \cos^{2} \left(\frac{\theta_{i+1}-\theta_{i}}{2} \right) + \left( \frac{1-\Gamma^{2}}{1+\Gamma^{2}} \right)^{2} \sin^{2} \left(\frac{\theta_{i+1}-\theta_{i}}{2} \right)},\\
    &\alpha_{i} = -\imath \left( \frac{2 \Gamma}{1+\Gamma^{2}} \right) \: \exp {-\imath \left(\frac{\theta_{i+1}-\theta_{i}}{2}\right)} \times \sin \left( \frac{\theta_{i+1}+\theta_{i}}{2} \right), \qquad \beta_{i}=\left( \frac{1-\Gamma^{2}}{1+\Gamma^{2}} \right),\\
    &\Delta_{i} = -\imath \left( \frac{2 \Gamma}{1+\Gamma^{2}} \right) \: \exp {-\imath \left(\frac{\theta_{i+1}-\theta_{i}}{2}\right)} \times \sin \left( \frac{\theta_{i+1} - \theta_{i}}{2} \right) \times \exp \left(\imath \theta_{i} \right).
  \end{aligned}
\end{equation}
\end{widetext}
The difference in Hamiltonian for conical and spiral configuration can be found by comparing Eq. (\ref{eq:conical-spin-aij}) and Eq. (\ref{eq:spiral-spin-aij}). In $a_{i}$ an additive term proportional to $\left(1-\Gamma^{2}\right)/\left(1+\Gamma^{2}\right)$ appear in the phase and the modulus part.  In $\alpha_{i}$ and $\beta_{i}$ only a pre-factor $2\Gamma/\left( 1 + \Gamma^{2} \right)$ appears. $\beta_{i}$ becomes non-zero. One should note that all these equations transforms to Eq. (\ref{eq:spiral-spin-aij}) when the spin lies on the $x-y$ plane ($\phi=\pi$, $\Gamma=1$). For constant spin modulation ($\theta_{i+1}-\theta_{i} \equiv \theta$ is constant) the effective Hamiltonian reads
\begin{equation}
  \label{eq:effectiv-ham-conical-spin}
  \begin{aligned}
    H_{\text{eff,con}}&=-\sum\limits_{i} \xi_{i+1} \bar{\xi}_{i} \: \tilde{t}_{i,\text{con}} - \left( \frac{2 \Gamma}{1+\Gamma^{2}} \right)\Delta \sin \frac{\theta}{2}\sum\limits_{i}\bar{\xi}_{i+1}\bar{\xi}_{i} + \text{H.c.} \\
    & \qquad + \left[ \mu + \left(\frac{1 - \Gamma^{2}}{1+\Gamma^{2}}   \right) B_{z}\right] \sum\limits_{i} \xi_{i}\bar{\xi}_{i},
  \end{aligned}
\end{equation}
where,
\begin{equation*}
  \begin{aligned}
    \tilde{t}_{i,\text{con}}
    =& t \:e ^{ \imath \atan \left[ \left( \frac{1- \Gamma^{2}}{1+\Gamma^{2}} \right) \tan \frac{\theta}{2} \right]}  \sqrt{\cos^{2} \frac{\theta}{2} + \left( \frac{1-\Gamma^{2}}{1+\Gamma^{2}} \right)^{2} \sin^{2} \frac{\theta}{2}}\\
    &\qquad + \left( \frac{2 \Gamma}{1+\Gamma^{2}} \right)\alpha \sin \left( \theta_{i} + \frac{\theta}{2} \right).
  \end{aligned}    
\end{equation*}
Qualitatively the only difference between Eq. (\ref{eq:effectiv-ham-conical-spin}) and Eq. (\ref{eq:effectiv-ham-spiral-spin}) is: (i) the hopping parameter, $\tilde{t}_{i,\text{con}}$ is complex valued, (ii) the effect of magnetic field is non-zero. One can see the emergence of pair density wave state when the phase of the complex $\tilde{t}_{i,\text{con}}$ is gauged out. Although investigation of the topological properties of pair density wave Hamiltonian is interesting in its own right \cite{cho-2014-topol-pair}, however, it is not the main aim of the current article. We will investigate thoroughly the phase diagram and topological properties of this Hamiltonian in another article.

\section{Conclusion}
\label{sec:conclusion}

In this work we systematically investigated the effect of the RSOC on the topological properties of the 1D nanowire in the \emph{e-e} strong correlation regime. The strong \emph{e-e} correlation allows for the fractionalization of the charge (holon) and spin (spinon) degrees of freedom, which we treat using the \emph{su(2|1)} path integral method. The resulting Hamiltonian in the presence of spiral spinon field (with modulation angle $\theta$) is given in Eq. (\ref{eq:effectiv-ham-spiral-spin}). It should be stressed that $\theta$ is a dynamic parameter (in the sense that it is a function of other system parameters); $\theta$ depends on spinon degree of freedom ($z_{i}$) --- $\theta$ is related to $z_{i}$ through Eqs. (\ref{eq:spin-config-spiral-xy}) and (\ref{eq:spin-cov-symb}). We find that for a minimal setup the magnetic field and RSOC are not necessary for topological phase to appear, provided \emph{e-e} correlation is strong enough. When $0<\alpha$ and $0<\theta<\pi$ by tuning $\mu$ the system can be driven into or out of the topological phase as shown in Fig. \ref{fig:seq-phase-diag}. It happens because non-zero $\alpha$ induces a quasi period --- through periodic electron hopping $\tilde{t}_{i}$, Eq. (\ref{eq:effectiv-ham-spiral-spin}) --- in the 1D effective Hamiltonian. It results in folding of the energy dispersion in the BZ and emergence of gaps at folded BZ boundary; explained schematically in Fig. \ref{fig:scheme-folding-of-BZ}. If the $\mu$ lies inside the energy gap then the topological phase is absent; if $\mu$ lies inside the band then topological present. Qualitatively, this situation is same as the original 1D Kitaev toy model \cite{kitaev_2001_UnpairedMajorana_Phys-Usp}. In the Kitaev toy model if the chemical potential lies inside the band ($-2<\mu/t<2$) then topological phase is present, however, if $\mu/t<-2$ or $2<\mu/t$ then the topological phase is absent. The only difference in our case is the occurrence of band gaps due to the folding of the BZ. We numerically validate the re-entrant topological properties as shown in Fig. \ref{fig:kitaev-chain-dep-on-mu} (Majorana energy spectrum) and Fig. \ref{fig:majorana-fm-ener-modes} (LDOS of Majorana fermions). In Fig. \ref{fig:seq-dynamic-theta} we calculate the $\theta$ corresponding the minimum free energy for $0<\mu/t<2$. We find that the system can be forced into or out of the topological phase just by tuning $\mu$. We propose heterostructure involving (quasi-)1D Moire structure, shown in Fig. \ref{fig:scheme-device-moire}, to investigate the predicted effect. We also investigated the behaviour of the system under an in-plane magnetic field in Sec. \ref{sec:contr-peri-hamilt}. It shows that the modulation angle can be controlled externally by the combination of the magnetic field and electric field. A heterostructure for investigation of the corresponding physics is proposed in Fig. \ref{fig:scheme-InAs}. The effect of conical spin structure has also been investigated in Sec. \ref{sec:conic-spin-struct}, which predicts the emergence of the pair-density wave states.

\section{Acknowledgements}
K.K.K wish to thank P. A. Maksimov for valuable discussion. K.K.K acknowledges the financial support from the JINR grant for young scientists and specialists, the Foundation for the Advancement of Theoretical Physics and Mathematics ”Basis” for grant \# 23-1-4-63-1. One of us, A.F., acknowledges financial support from the MEC, CNPq (Brazil) and from the Simons Foundation (USA).

\appendix

\section{Extended topological phase diagram}
\begin{figure*}[tbh]
  \centering
  \includegraphics[width=0.95\textwidth]{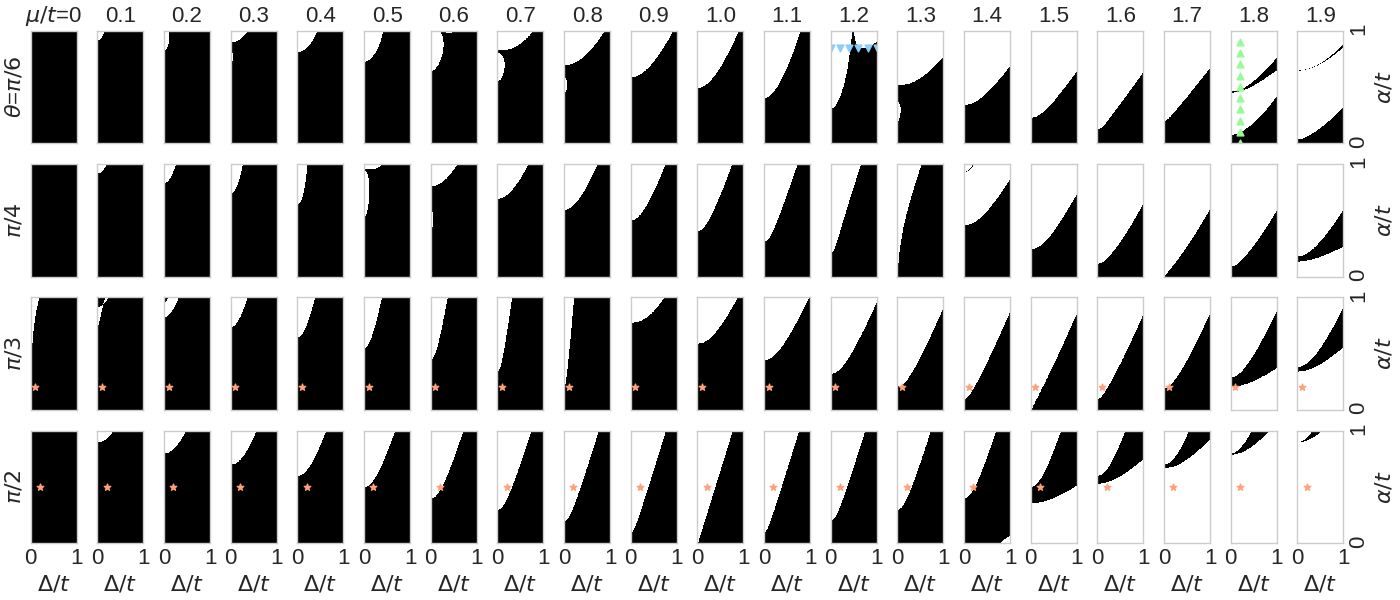}
  \caption{Color online: Dependence of $\mathbb{Z}_{2}$ topological invariant on chemical potential ($\mu$), SC gap ($\Delta$), and RSOC ($\alpha$) and spiral spinon texture with phase modulation angle $\theta$. The topologically trivial (non-trivial) phase is shown in white (black). In the figure the columns represent the chemical potential $\mu/t$ (increasing from left to right). The rows represent the phase of the spinon texture $\theta=\pi/6,\pi/4,\pi/3,\pi/2$. $x$-axis of each sub-figure represents the SC gap $\Delta/t \in \left( 0,1 \right)$ and y-axis represent the RSOC $\alpha/t \in \left( 0,1 \right)$. For $\theta=\pi/2$ (last row, $\alpha/t=0.5$, $\Delta/t=0.2$) and $\theta=\pi/3$ (second last row, $\alpha/t=0.2$, $\Delta/t=0.1$) the points marked by star (red) represent the re-entrant topological transition --- from topologically non-trivial (black) to trivial (white), to again non-trivial phase (black) --- as $\mu/t$ is varied. The same transition is also shown as $\Delta$ (first row, $\theta=\pi/6, \mu/t=1.2$, blue down-triangle) and $\alpha$ (first row, $\theta=\pi/6, \mu/t=1.8$, green up-triangle) are varied.}
  \label{fig:seq-phase-diag}
\end{figure*}
As mentioned in Sec. \ref{sec:phase-diagram} the topological invariant $Q$ is related to $A_{k}$, which is a function of parameters $\mu$, $\theta$, $\Delta$ and $\alpha$. Therefore an interesting topological phase diagram can be found by tuning accordingly these parameters. In Fig. \ref{fig:seq-phase-diag} we plot the extended phase diagram of the  dependence of $Q$ on these parameters, specifically, we plot the topological invariant $Q$ on the $\Delta$--$\alpha$ parameter space for different combination of $\mu$ and $\theta$. It can be observed that at constant $\theta$ (every row) and increasing $\mu$ (from left to right) the area of the topologically non-trivial phase (black) decreases. We also observe that, in each row the trivial phase starts to seep in from higher $\alpha$ and lower $\Delta$ as $\mu/t$ is increased. For example, when $\theta=\pi/6$ (first row) and $\mu/t=0.01$ the whole $\Delta$--$\alpha$ parameter space has a non-trivial topological phase ($Q=-1$, black). When $\mu/t=0.1,0.2$ the trivial phase ($Q=1$, white) starts to seep in from the upper left corner. With a further increase in $\mu/t$ the area of topological phase increases. This can be understood by keeping in mind the energy spectrum of the archetypal Kitaev chain and by observing the momentum space energy spectrum of the Hamiltonian, Eq. (\ref{eq:ham-mom}) \cite{kitaev_2001_UnpairedMajorana_Phys-Usp,alicea_2012_NewDirections_RepProgPhys} \footnote{For an archetypal Kitaev chain the energy spectrum as a cosine dependence \cite{alicea_2012_NewDirections_RepProgPhys}. The topological boundary corresponds to $\left| \mu \right| \leq 2t$; physically it corresponds to the maximum ($2t$) and minimum ($-2t$) of the energy band.}. Although Eq. (\ref{eq:effectiv-ham-spiral-spin}) is analogous to the Hamiltonian of the archetypal Kitaev chain in real space, but in momentum space the energy spectrum differs from each other due to presence of $\theta$. Effect of inclusion of $\theta$ in the Hamiltonian are of two folds. First, with the increase in $\theta$ the energy dispersion becomes flatter (hopping is proportional to $t \cos (\theta/2)$, hence the width of band decreases), therefore, the electrons become more localized. Second, when $\alpha \neq 0$, depending on periodicity of $\tilde{t}_{i}$ (periodicity depends on $\theta$) corresponding number of band gaps appear in the energy spectrum; when $\alpha=0$ only a single band is present as $\tilde{t}_{i}$ loses its periodicity. In Fig. \ref{fig:ener-spect-alpha-dep} we plot the energy spectrum for $\theta=\pi/3$ and $\alpha/t=0.2,0.4,0.6,0.8$. We see that six energy bands are present for each $\alpha$; it was expected as periodicity of $\tilde{t}_{i}$ in this case is sixfold. Besides, we also observe that with the increase in $\alpha$ the energy gap increases, and bands become flatter. Now the reason for the onset of the trivial phase from the values $\alpha \approx 1, \Delta \approx 0$ (upper left corner of each row) in Fig. \ref{fig:seq-phase-diag} is clear. At high $\alpha$ the bands are flatter compared to the ones with low $\alpha$. Therefore, when for higher $\alpha$ we gradually increase $\mu$ the maximum of the band is reached first, compared to the lower $\alpha$ case. For example in Fig. \ref{fig:ener-spect-alpha-dep} when $\alpha/t=0.8$ the $\mu/t=0.5$ level already lies inside the band gap between the first and the second bands (the bands are counted from $\mu/t=0$), whereas, for $\alpha/t=0.2$ it lies inside the first band. Therefore in the Fig. \ref{fig:seq-phase-diag} for $\theta=\pi/3$, $\Delta/t \approx 0$ and $\mu/t=0.5$ the system is already in trivial topological phase for $\alpha/t=0.8$, although, for $\alpha/t=0.2$ the topological non-trivial phase is still present in the system.

Another interesting feature we observe in Fig. \ref{fig:seq-phase-diag}, is the multiple transition from topologically trivial to non-trivial phase (and vice-versa) as the parameters are varied (we call it ``\emph{re-entrant}'' topological phase transition). For example the system for $\theta=\pi/3$, $\alpha/t=0.2$, $\Delta/t=0.1$ (red star, second last row in Fig. \ref{fig:seq-phase-diag}) goes through phase transitions three times as $\mu$ increases: (i) at $\mu/t=1.3$ from non-trivial to trivial phase, (ii) at $\mu/t=1.8$ from trivial to non-trivial phase, (iii) at $\mu/t=1.9$ from non-trivial to trivial phase. This can be qualitatively understood from the energy spectrum diagram of Fig. \ref{fig:ener-spect-alpha-dep}. In Fig. \ref{fig:ener-spect-alpha-dep} for $\alpha/t=0.2$ three bands are present. As $\mu$ is increased the system goes through all these bands. The transition from non-trivial (trivial) to trivial (non-trivial) phase occurs when $\mu$ goes from inside the band (band gap) into the band gap (band). In the band gap there is not electronic states to generate MFs, hence, the phase is trivial. This is analogous to the behavior of the archetypal Kitaev chain \cite{kitaev_2001_UnpairedMajorana_Phys-Usp}. In Kitaev chain due to absence of the periodicity only single band is present. When the $\mu$ lies inside the band $-2<\mu/t<2$ then non-trivial phase is present. However, when $\mu$ lies outside this range the phase is trivial. In our case due to the periodic nature of the Hamiltonian (when $\alpha >0 $) and when the BZ gets folded ($N$ times if periodicity is $N$). It results in the band gap at the boundary of the folded BZ. If the $\mu$ lies in the band gap then the phase is trivial; if the $\mu$ lies inside the band then phase is non-trivial. In fact, not only $\mu$, but also, by varying $\Delta$ (blue down triangles in first row of Fig. \ref{fig:seq-phase-diag}, $\theta=\pi/6$, $\mu/t=1.2$) and $\alpha$ (green up triangles in first row of Fig. \ref{fig:seq-phase-diag}, $\theta=\pi/6$, $\mu/t=1.8$) the re-entrant topological phase transition can be achieved.

\section{Coherent symbol representation of the Hubbard operators}
\label{sec:su21}

The $su(2|1)$ superalgebra can be thought of as the simplest possible extension of the conventional spin $su(2)$ algebra to incorporate fermionic degrees of freedom \cite{ferraz_2011_EffectiveAction_NuclearPhysicsB}. Namely, the  bosonic sector of the $su(2|1)$ consists of three bosonic superspin operators,
\begin{equation}
  Q^{+}=X^{\uparrow\downarrow},\quad Q^{-}=X^{\downarrow\uparrow},\quad
  Q^{z}=\frac{1}{2}(X^{\uparrow\uparrow}-X^{\downarrow\downarrow})
  \label{qx}\end{equation}
closed into $su(2)$, and a bosonic operator $X^{00}$ that generates a $u(1)$ factor of the maximal even subalgebra $su(2)\times u(1)$ of $su(2|1)$. The fermionic sector is constructed out of four operators $X^{\sigma 0}, X^{0\sigma}$ that transform in a spinor representation of $su(2)$.

A coherent-state symbol for a Hubbard operator reads
\begin{equation}
  X \equiv \langle z,\xi|X|z,\xi\rangle.
  \label{symbol}\end{equation}
The states $|z,\xi\rangle$ are the same as defined in Eq. (\ref{eq:cs-symbol-su-1}). Using Eq. (\ref{symbol}) and (\ref{eq:cs-symbol-su-1}) the fermionic sector operators are derived as
\begin{eqnarray}
  \label{x}
  X^{0 \downarrow}&=&-\frac{z\bar\xi}{1+|z|^2},\quad
                      X^{\downarrow 0}=-\frac{\bar
                      z\xi}{1+|z|^2},\nonumber\\
  X^{0\uparrow}&=&-\frac{\bar\xi}{1+|z|^2}, \quad X^{\uparrow
                   0}=-\frac{
                   \xi}{1+|z|^2}.
\end{eqnarray}
Similarly the bosonic sector operators, as defined in Eq. (\ref{qx}), are derived as
\begin{eqnarray}
  \label{1.3}
  Q^{+}&=&S^{+}_{cov}\left(1-X^{00}_{cov}\right),\quad Q^{-}=S^{-}
           \left(1-X^{00}\right),\nonumber\\
  Q^z&=&S^z
         \left(1-X^{00}\right).
\end{eqnarray}
Here $X^{0,0}$ is the covariant symbol of the hole number operator; it is reads
\begin{equation}
  \label{1.3a}  
  X^{00}=\frac{\bar\xi\xi}{1+|z|^2}.
\end{equation}
Similar to coherent state symbol of Hubbard operators, the coherent state symbols of $su(2)$ generators are evaluated to be:
\begin{equation}
  \label{eq:su-2-generators}
   S \equiv \langle z|S|z\rangle.
 \end{equation}
Remember it acts only on the spinon ($z$) degrees of freedom. Explicitly, the coherent symbols can be calculated by using Eqs. (\ref{eq:cs-symbol-su-1}) and (\ref{eq:su-2-generators}):
\begin{eqnarray}
  \label{1.3b}  
  S^{+}=\frac{z}{1+|z|^2},
  S^{-}=\frac{\bar z}{1+|z|^2},
  S^z=\frac{1}{2}\left(\frac{1-|z|^2}{1+|z|^2}\right)
\end{eqnarray}
For a compact simple (super)algebra these symbols are in one-to-one correspondence with the algebra generators. This is the case for both the $su(2|1)$ and $su(2)$ algebras. Using Eqs. (\ref{x}) and making a change of variables
\begin{equation}
  \label{change}  
  z\to z,\quad \xi\to\xi\sqrt{1+|z|^2},
\end{equation}
we arrive at Hamiltonian Eq. \ref{eq:effectiv-ham-trans}.

\end{document}